\documentclass[12pt]{iopart}

\expandafter\let\csname equation*\endcsname\relax

\expandafter\let\csname endequation*\endcsname\relax

\usepackage{amsmath}
\usepackage{amssymb}
\usepackage{caption}
\usepackage{bm}
\usepackage{graphicx}
\usepackage{hyperref}
\usepackage{array}
\usepackage{color}
\usepackage{subcaption,mwe}
\usepackage[export]{adjustbox}
\usepackage[toc,page]{appendix}
\usepackage[url=false, doi=false, backend=biber, style=numeric]{biblatex}
\DeclareUnicodeCharacter{0229}{e}
\DeclareUnicodeCharacter{03F5}{epsilon}
\DeclareUnicodeCharacter{03B5}{epsilon}
\DeclareUnicodeCharacter{03C4}{tau}

\newcommand{\be}{\begin{equation}}
\newcommand{\ee}{\end{equation}}

\begin{document}

\title[Spontaneous stochasticity in a 3d Weierstrass-ABC flow]{Spontaneous stochasticity in a $3d$ Weierstrass-ABC  flow}

\author{Antoine Barlet, Adam Cheminet, Bérengère Dubrulle}
\address{SPEC/IRAMIS/DSM, CEA, CNRS, University Paris-Saclay, France}
\ead{antoine.barlet@inria.fr}
\author{Alexei A. Mailybaev}
\address{Instituto de Matem\'atica Pura e Aplicada – IMPA, Rio de Janeiro, Brazil}

\vspace{10pt}
\begin{indented}
\item[]August 2017
\end{indented}

\begin{abstract}

Chaotic systems are characterised by exponential separation between close-by trajectories, which in particular leads to deterministic unpredictability over an infinite time-window. It is now believed, that such butterfly effect is not fully relevant to account for the type of randomness observed in turbulence. For example, tracers in homogeneous isotropic flows are observed to separate algebraically, following a universal cubic growth, independent from the initial separation. This regime, known as Richardon’s regime, suggests that at the level of trajectories, and unlike in chaos theory, randomness may in fact emerge in finite-time. This phenomenon called ‘spontaneous stochasticity’ originates from the singular nature of the underlying dynamics, and provides a candidate framework for turbulent randomness and transport. While spontaneous stochasticity has been mathematically formalised in simplified turbulence models, a precise and systematic tool for quantifying the various facets of this phenomenon is to this day missing. In particular, it is still unclear whether chaos is important for that behaviour to appear.

In this paper we introduce a 3d rough flow that can be tuned to present Lagrangian chaos. The flow is inspired by the Weierstrass function and is entitled 'the WABC model'. After analysing its properties, we define what is spontaneous stochasticity in this context. The provided formal definition is then adapted to better suit for numerical analysis. We present the results from Monte-Carlo simulations of Lagrangian particles in this flow. Within the numerical precision, we quantitatively observe the appearance of spontaneous stochasticity in this model. We investigate the influence of noise type and find that the observed spontaneous stochasticity does not depend on the chosen stochastic regularisations.

\end{abstract}

%
%
%
%
%

\section{Introduction}
Since the work of Poincaré, we know that smooth systems can display stochasticity in the long time while being totally deterministic. They are called chaotic systems and are such that two trajectories in phase space, initially separated by a small $\epsilon$, separate at an exponential rate. The rate of separation is the so-called Lyapunov exponent. In such chaotic flows, the two trajectories become undistinguishable from each other in the limit of a vanishing initial separation $\epsilon$.
While scientists believed for a long time that turbulence and its consequences (like weather) are chaotic, it is now clear that the unpredictability in developed turbulence is conceptually different. Studies of the Navier-Stokes turbulence reveal that two flows starting from two initial conditions separated by a distance $\epsilon$ remain distinct in the limit of $\epsilon \to 0$, provided that this limit is taken together with the increasing Reynolds number, $\mathrm{Re} \to \infty$ \cite{leith1972predictability,ruelle1979microscopic,eyink1996turbulence,thalabard_butterfly_2020}. This confirms Lorenz' 1969 prediction \cite{lorenz_predictability_1969}, according to which  multi-scale systems, like fluids, can remain unpredictable at finite times even in the limit of infinitely small initial uncertainty.

Such behaviour can also be observed from the Lagrangian point of view. Richardson indeed showed \cite{richardson_atmospheric_1926}, that two trajectories initially separated in the real space by a small $\epsilon$ diverge from each other at a $t^3$ rate, independently of $\epsilon$. This explosive separation has been observed in numerical simulations \cite{thalabard_turbulent_2014, scatamacchia_extreme_2012, bitane_time_2012} and for atmospheric data, but its experimental confirmation is still uncertain \cite{weiss_lagrangian_2024}. Its analysis has been done in toy models of turbulence, starting from Kraichnan model, where the advecting velocity field is replaced by a fractal Brownian motion white in time. In such model, deterministic Lagrangian trajectories correspond to a situation where all the particles stick together, while stochastic trajectories correspond to explosive separation of trajectories \cite{vanden_eijnden_generalized_2000,falkovich_particles_2001,kupiainen2003nondeterministic}. The transition between deterministic and stochastic trajectories resembles a phase transition. While being first called 'intrinsic stochasticity', this transition was called 'spontaneous stochasticity' \cite{falkovich_particles_2001}, in a direct allusion to symmetry breaking in phase transition theory.

It appears from the analysis of simplified models that irregularities are necessary for such behaviour to appear~\cite{eyink_renormalization_2020,drivas_statistical_2024}. A stochastic regularisation is also needed to solve the dynamics and properly converge towards this behaviour \cite{mailybaev2016spontaneously,mailybaev2017toward,biferale2018rayleigh,mailybaev_spontaneously_2023,bandak_spontaneous_2024}. In the end, spontaneous stochasticity corresponds to the paradoxical remaining stochasticity in the limit of vanishing regularisation. A general recipe to obtain such phenomenon is still under investigation \cite{mailybaev2023spontaneous,mailybaev2024rg} with still various open questions. Especially, it is not clear if chaos is particularly needed to get this regime. In that respect, it would be interesting to have at our disposal a tunable 3d toy model, that shares many similarities with turbulent flows, and that can be used to test ideas and intuitions. The goal of the present paper is to build such a model, the 'WABC flow'.

This model is  a $3d$ irregular model built from stationary Euler solutions called the ABC flows. We use these flows as building blocks in a multi-scale velocity field, in the prospect of observing Lagrangian spontaneous stochasticity. The first part of this paper is dedicated to introducing the model and presenting its general properties. We particularly investigate its symmetries and intermittency. We also define anomalous dissipation based on arguments presented for other close models. We analyse their statistics and show surprising non-Gaussian results.

The condition of observing spontaneous stochasticity is then introduced in a second part. We there define the stochastic regularisations that we use. A general definition for spontaneous stochasticity is then given in the context of the WABC model.  Since the full probability distributions are not accessible for numerical investigation, we study the spontaneous stochasticity by computing their finite-dimentional marginals. We finally discuss the numerical integration of Lagrangian trajectories using Monte-Carlo simulations.

Results from  numerical simulation are finally presented in a third and last part. Within numerical precision, we indeed observe spontaneous stochasticity qualitatively and quantitatively. We also compare two regularisations and show that they lead to the same results in the limit of vanishing noise. We finally discuss the potential extension of those results for other simulation parameters. Since ABC flows -- and by extension the WABC flow -- display Lagrangian chaos for some well-chosen parameters, the WABC model should be well suited to study the influence of chaos on the appearance of spontaneous stochasticity.

\section{The WABC model}
\subsection{Definitions}
\subsubsection{The ABC model --}

We first introduce the ABC flow, a flow that was created by V. Arnold in 1965 \cite{arnold_sur_1965} following the works of E. Beltrami and S. Childress. It is defined as 
\be
\bm U(\bm x) = \begin{cases}
	A \sin(z) + C \cos(y) \\
	B \sin(x) + A \cos(z) \\
	C \sin(y) + B \cos(x),
\end{cases}
\ee
where $A$, $B$ and $C$ are constants. This flow solves the incompressible stationary Euler equations on the torus $[0, 2 \pi]^3$ and has the special property of having vorticity $\bm \omega$ equal to velocity (a special case of Beltrami property where $\bm \omega \propto \bm U$). This solution of Euler equation has the advantage of being simple and analytical.

As pointed out by Dombre et al. \cite{dombre_chaotic_1986}, the following system of equations for Lagrangian particles
\be
\begin{cases}
	\dot x = A \sin(z) + C \cos(y) \\
	\dot y = B \sin(x) + A \cos(z) \\
	\dot z = C \sin(y) + B \cos(x),
\end{cases}
\ee
remains unchanged under the following symmetries
\be
\begin{cases}
	S_1: \; x \to x', \; y \to \pi - y', \; z \to -z', \; t \to -t' \\
	S_2: \; x \to -x', \; y \to y', \; z \to \pi - z', \; t \to -t' \\
	S_3: \; x \to \pi - x', \; y \to - y', \; z \to z', \; t \to -t'.
\end{cases}
\ee
A flow invariant under those transformations is considered to be time-reversible following Birkhoff's definition. It was also shown to display Lagrangian chaos for some specific $A$, $B$ and $C$ coefficients, as mentioned by Hénon \cite{henon_sur_1966}. We however do not intend to study the effect of chaos in this paper. We nevertheless choose to have $A = 1.732$, $B = 1.314$ and $C = 1$ which appears to be a case that builds Lagrangian chaos.

\subsubsection{Building the WABC model --}
The assumption of having a smooth dynamical setup has to be broken to build a spontaneously stochastic model: we need irregularities. We construct a $3d$ 'Weierstrass' velocity field with irregularities in space. The elementary brick is the ABC flow, having similarities with other models \cite{elliott_pair_1996, eyink_suppression_2013}. We call this model the 'Weierstrass-Arnold-Beltrami-Childress' or 'WABC' and define it as:
\be \label{WABSeq}
\bm u_W(\bm x) = \sum_{i=1}^{\infty} \frac{\omega_i}{k_i} \, \bm U(k_i \bm x).
\ee
where $k_i = \lambda^i$ and $\omega_i = \lambda^{(1-h) i}$, having $\lambda > 1$. To keep $\bm u_W$ defined on the torus $[0, 2 \pi]^3$, we choose to set $\lambda = 2$.

\begin{figure}
	\centering
	\begin{subfigure}[t]{0.45\textwidth}
		\captionsetup{width=0.90\textwidth,margin={0mm,0.4\textwidth}}
		\centering
		\includegraphics[width=10cm, right]{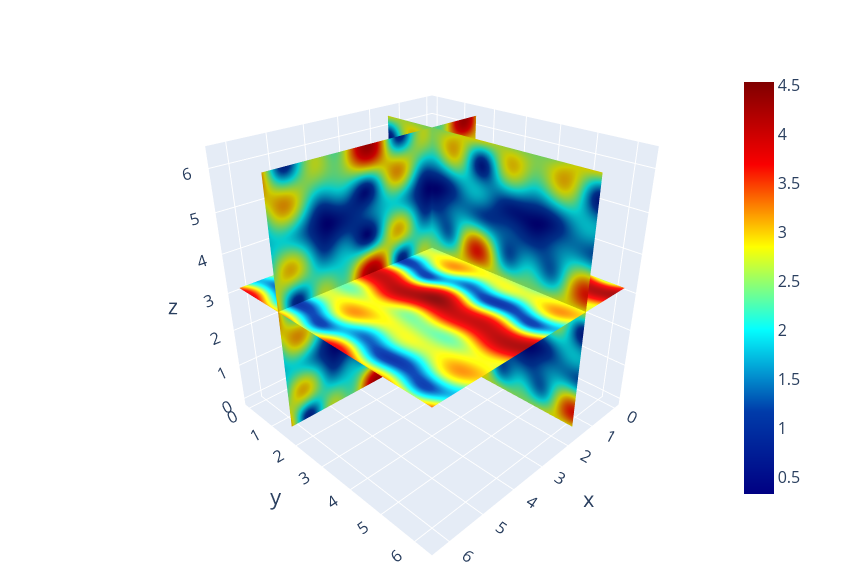}
		\caption{$N = 2$}
	\end{subfigure}
	\begin{subfigure}[t]{0.45\textwidth}
		\captionsetup{width=0.90\textwidth,margin={0.4\textwidth,0mm}}
		\centering
		\includegraphics[width=10cm, left]{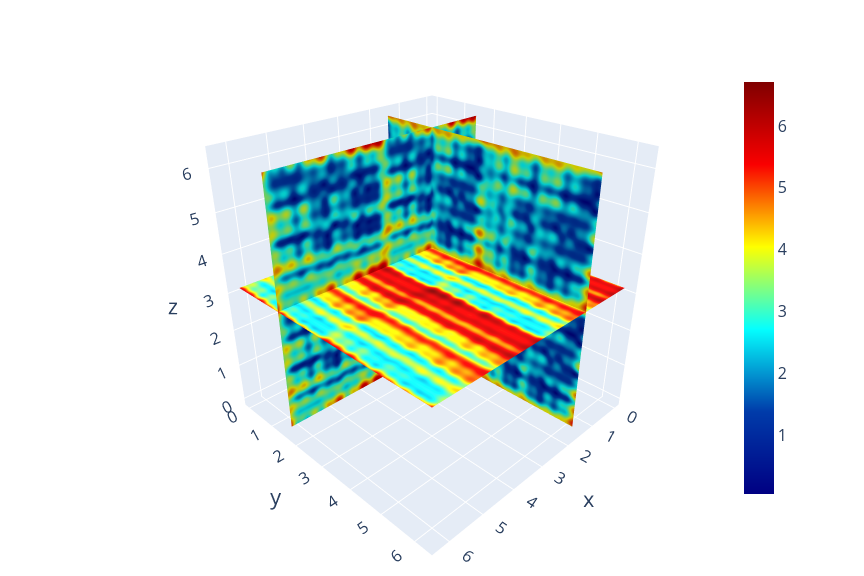}
		\caption{$N = 4$}
	\end{subfigure}
	\begin{subfigure}[t]{0.45\textwidth}
		\centering
		\includegraphics[width=10cm, center]{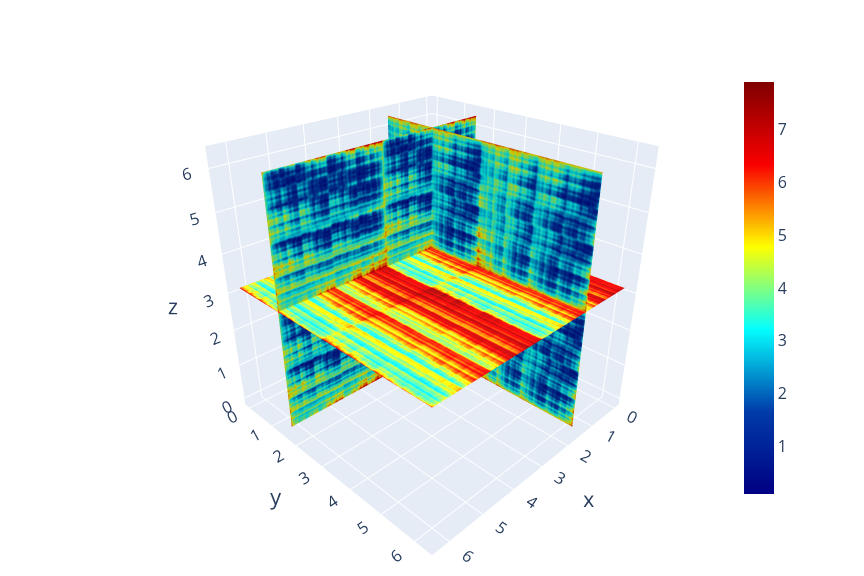}
		\caption{$N = 6$}
	\end{subfigure}
	\caption{$||\bm u_W||_2$ slices of the WABC flow for increasing number of modes $N$.}
	\label{fig:SlicesWABC}
\end{figure}

The above velocity field is bounded by the exponential series $\sum_i^{\infty} \lambda^{-h i}$ which converges for $h > 0$. This stationary $3d$ flow is a succession of layers each representing a smaller version of the ABC flow (see Figure \ref{fig:SlicesWABC}).

We note that this flow does not solve Euler equations, even in its weak formulation, since the non-linear term $(\bm u_W \cdot \bm \nabla) \bm u_W$ breaks the superposition property. Also, since it is deeply connected to the Weierstrass function, one can show that the WABC flow is h-Hölder continuous. As a result, almost all spatial derivatives are undefined for $h < 1$. This leads to diverging vorticity in that range of Hölder exponents. This flow is still incompressible for any $h$ by linearity of divergence,
\be
\bm \nabla \cdot \bm u_W = \sum_{i=1}^{\infty} \frac{\omega_i}{k_i} \, \bm \nabla \cdot \bm U (k_i \bm x) = 0.
\ee

\subsection{General properties}
\subsubsection{Symmetries --}
Symmetries $S_1$, $S_2$ and $S_3$ are broken under the addition of modes of ABC flows. Therefore, the WABC flow cannot be seen as time reversible contrary to the original ABC model. We finally note that due to contributions of all modes, Beltrami property is broken for any $h \neq 1$. 

\subsubsection{Self-similarity --}
\begin{figure}
	\centering
	\begin{subfigure}[t]{0.45\textwidth}
		\captionsetup{width=0.90\textwidth,margin={0mm,0.3\textwidth}}
		\centering
		\includegraphics[width=9cm, right]{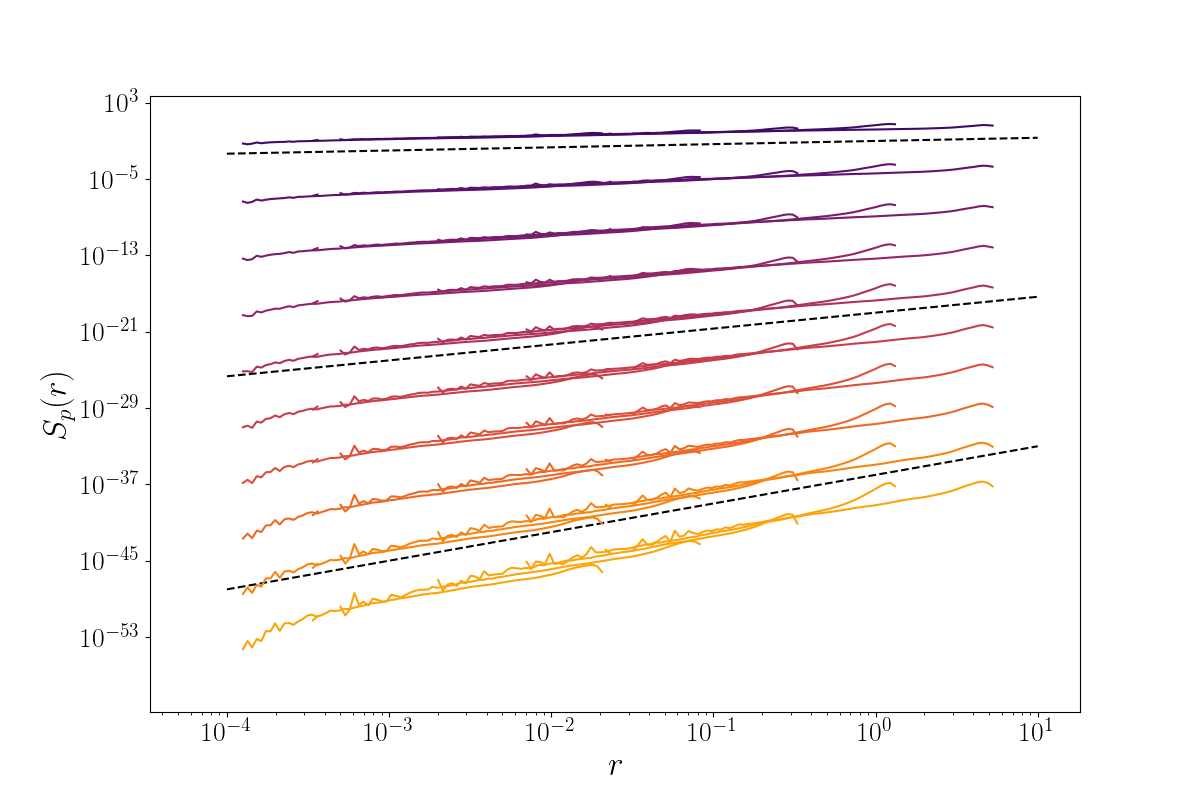}
		\caption{}
	\end{subfigure}
	\begin{subfigure}[t]{0.45\textwidth}
		\captionsetup{width=0.90\textwidth,margin={0.3\textwidth,0mm}}
		\centering
		\includegraphics[width=9cm, left]{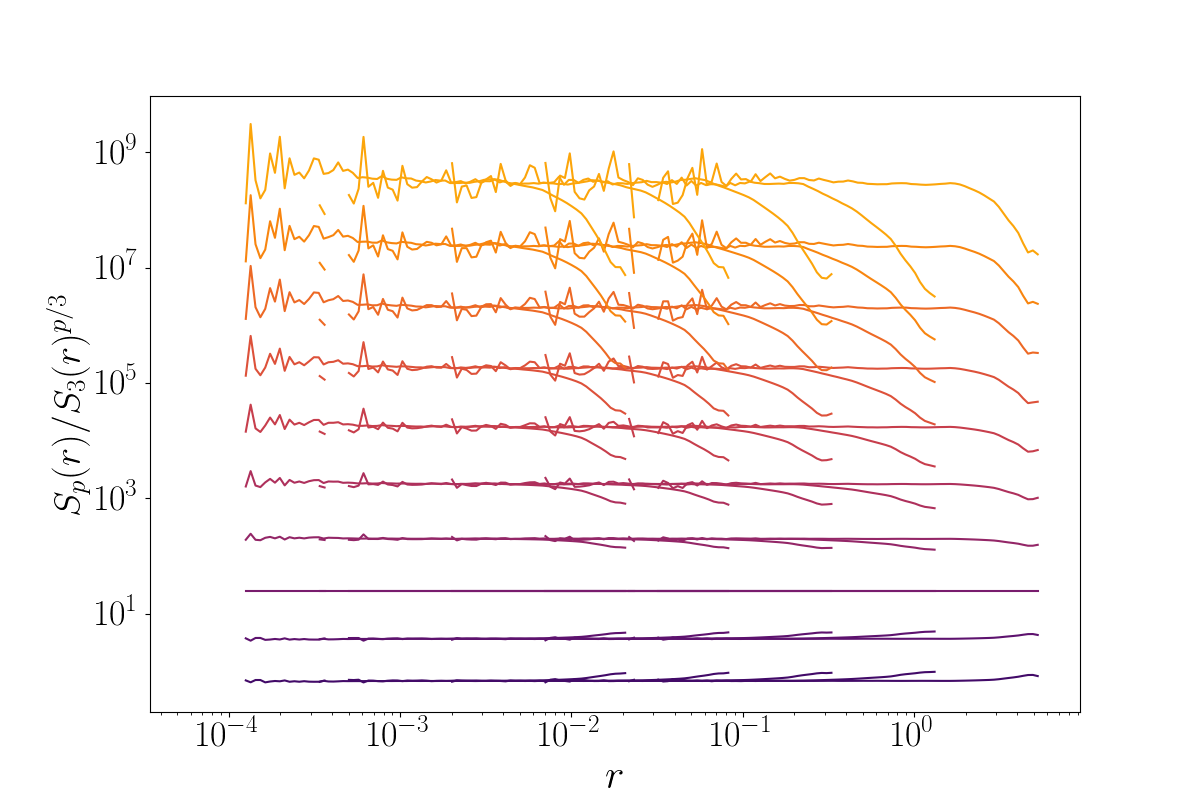}
		\caption{}
	\end{subfigure}
	
	\caption{Eulerian structure functions $S_p(r) = \langle |\delta u_{\parallel}|^p \rangle$ as a function of distance $r$, not normalised (a) and normalised by $S_3(r)^{p/3}$ (b) for the WABC flow with $h = 1/3$ and $N = 14$. Increasing $p$ are displayed by colours, ranging from $p=1$ (dark purple) to $p=10$ (light orange). The dotted lines have the slope $p/3$.}
	\label{fig:StrucFunc}
\end{figure}

Even though the full self-similarity is broken, the WABC model still has a partial self-similarity property since: 
\be 
\bm u_W(\bm x) = \lambda^{-h} \bm u_W(\lambda \bm x) + \lambda^{-h} \bm U(\lambda \bm x),
\ee
which suggests that we could find self-similar structures. This can be studied using the longitudinal velocity increments over a distance $\bm r$, defined as  $\delta u_{\parallel}(\mathbf{x},\bm r)= \big[\bm u_W(\bm x+\bm r)-\bm u_W(\bm x)\big] \cdot \frac{\mathbf{r}}{|\bm r|}$. The Eulerian structure functions $S_p(r) = \langle |\delta u_{\parallel}|^p \rangle$ are obtained by averaging over the space $\bm x$ and the directions $\bm r/r$ at a fixed distance $r = |\bm r|$. Their numerical computation for the WABC flow with $h = 1/3$ is presented in Figure \ref{fig:StrucFunc}. We see that all structure functions behave as power laws with exponent $\zeta_p = p/3$ for scales larger than the cut-off. This shows that the flow is mono-fractal, with exponent $h = 1/3$.

\subsection{Anomalous dissipation}
\subsubsection{Definition --}
The WABC flow is not a solution to Euler or Navier-Stokes equations, but we can still define the notion of local energy transfers formally. Rather than using Eyink's ideas of local energy transfer between layers of Weierstrass-type flow \cite{eyink_energy_1994}, we prefer to stick to the local energy transfer term introduced by Duchon and Robert (2000) for the Navier-Stokes equations \cite{duchon_inertial_2000}. It is defined as: 
\be \label{eq:DefDR}
\Pi^\ell = \frac{1}{4} \int \nabla \psi^\ell(\bm r) \cdot \delta_{\bm r} \bm u \left|\left|\delta_{\bm r} \bm u \right|\right|_2^2 \, d \bm r,
\ee
where $\delta_{\bm r} \bm u = \bm u_W(\bm x+\bm r)-\bm u_W(\bm x)$ are the velocity increments and  $\psi^\ell$ is a mollifying function, which is  positive, even and  infinitely differentiable with compact support. Since the WABC flow is $h-$Hölder continuous, it is straightforward to show that:
\be
\Pi^\ell = O(\ell^{3h - 1}).
\ee
Therefore, for any $h > 1/3$, we expect $\Pi^\ell$ to vanish as $\ell\to 0$.  In the opposite case, $h < 1/3$, this term is unbounded. In case $h = 1/3$ it reaches a finite value and corresponds to the "dissipation anomaly" observed in turbulent flows, i.e. an inertial dissipation independent of viscosity \cite{onsager_statistical_1949, dubrulle_beyond_2019}.

\begin{figure}
	\centering
	\includegraphics[width=12cm]{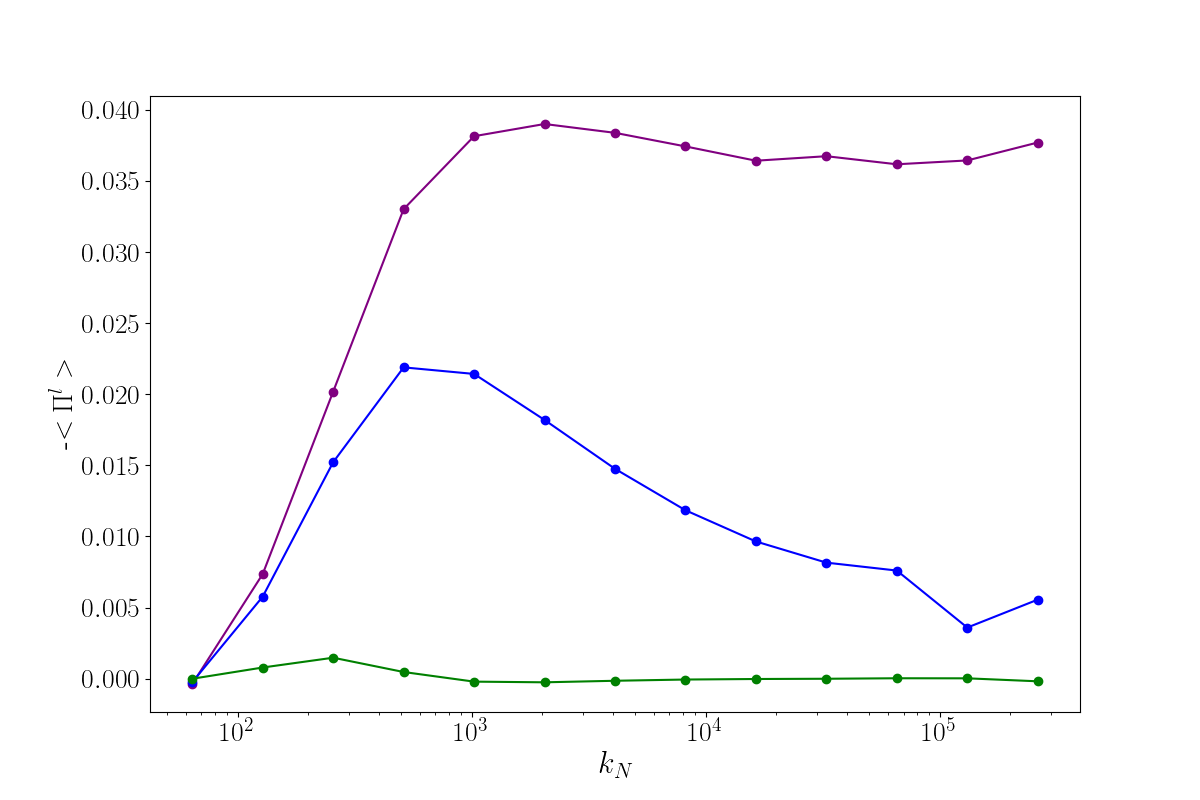}
	
	\caption{Evolution of mean Duchon-Robert coefficient  $\left< \Pi^\ell \right>$ as a function of $k_N$ for the WABC model when: $h = \frac{1}{3}$ (purple), $h = \frac{2}{5}$ (blue) and $h = 1$ (green).}
	\label{fig:AnomDissWABC}
\end{figure}

\begin{figure}
	\centering
	\begin{subfigure}[t]{0.45\textwidth}
		\captionsetup{width=0.90\textwidth,margin={0.30\textwidth,0mm}}
		\centering
		\includegraphics[width=12cm, center]{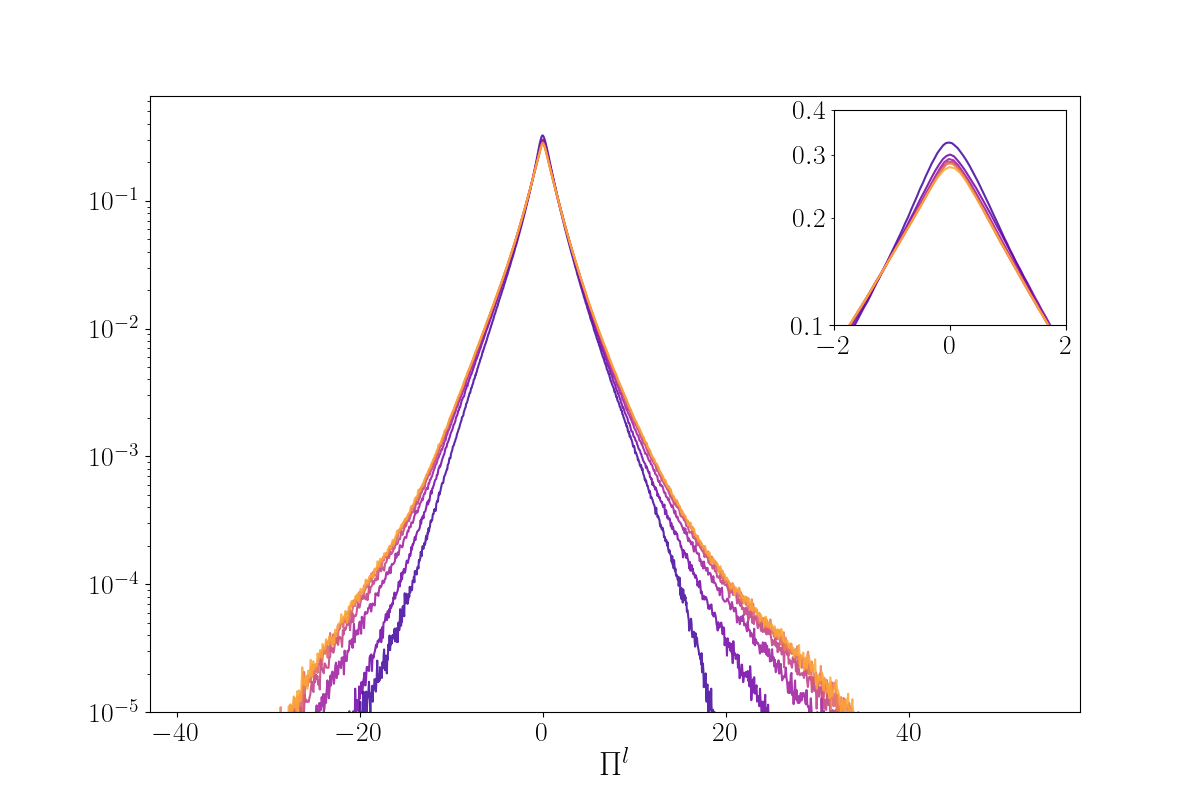}
	\end{subfigure}
	
	\caption{Probability distribution of Duchon-Robert coefficient $\Pi^\ell$ as a function of $N$, ranging from $N=6$ (dark purple) to $N = 18$ (yellow) for the WABC flow with $h=1/3$.}
	\label{fig:PdfEpsWABC}
\end{figure}

For the numerical analysis, we truncate the WABC flow as
\be \label{WABSeqN}
\bm u_W^{(N)}(\bm x) = \sum_{i=1}^{N} \frac{\omega_i}{k_i} \, \bm U(k_i \bm x),
\ee
where the cutoff parameter $N$ denotes the number of modes. 
We evaluate the global statistics followed by $\Pi^\ell$ for different scales $\ell$ and number of modes $N$. We scale $\ell$ with $N$ by setting $\ell = 2.8 k_N^{-1}$. We show in Figure \ref{fig:AnomDissWABC} the evolution of $\langle \Pi^\ell\rangle$ as a function of $k_N$ for $h = 1/3$, $h = 2/5$ and $h = 1$. As expected, for $h > 1/3$, the curves  vanish in the limit of $N \to \infty$. For $h = 1/3$ the curve reaches a plateau, mimicking the presence of inertial dissipation. Note that the sign of $\langle \Pi^\ell \rangle$ can be reversed by the transformation $\bm u_w\to -\bm u_w$.
Figure \ref{fig:PdfEpsWABC} shows the evolution of $p(\Pi^l)$ as a function of $N$: the probability distributions become more peaked when converging towards the inviscid limit. When $h = 1/3$, the distributions collapse onto a unique non-Gaussian distribution. 

\section{Spontaneous stochasticity in the WABC model}

\subsection{Stochastic regularisations}
To solve the Lagrangian dynamics, we consider two types of stochastic regularisations depending on the cutoff parameter $N$. Note that a deterministic regularisation does not allow for spontaneous stochasticity to appear and it would lead to ill-defined limits  \cite{drivas_statistical_2024}. 

In the first type of regularisation, positions of the particles $\bm x^{(N)} (t)$ are governed by the Langevin-type stochastic differential equation
\be \label{eq:LangWABC}
	d \bm x^{(N)} = \bm u_W^{(N)}\left(\bm x^{(N)}\right) dt 
    +  \sqrt{2 \kappa_N} \, d\bm {W}, \quad
    \bm x^{(N)} (0) = \bm x_0.
\ee
where $d\bm {W}$ is a three-dimensional white-noise term.  The regularisation (diffusion) parameter $\kappa_N$ must be chosen such that $\kappa_N \to 0$ as $N \to \infty$. Then the regularised system (\ref{eq:LangWABC}) converges formally to the initial-value problem for the WABC flow (\ref{WABSeq}). The initial condition $\bm x_0$ is considered to be deterministic. We will refer to this regularisation as 'Langevin-WABC'.

In the second type of regularisation, positions of the particles $\bm x^{(N)} (t)$ are governed by the (deterministic and Lipschitz continuous) differential equation obtained by a simple truncation of the WABC flow (\ref{WABSeqN}). The stochasticity is introduced in the initial condition. The resulting system takes the form
\be \label{eq:LangWABC2}
	\frac{d \bm x^{(N)}}{dt} = \bm u_W^{(N)}\left(\bm x^{(N)}\right), \quad
    \bm x^{(N)} (0) = \bm x_0+\eta_N \bm w,
\ee
where $\bm w$ is a random variable uniformly distributed in a three-dimensional unit ball. The regularisation coefficients must be chosen such that $\eta_N \to 0$ as $N \to \infty$. Then the regularised system (\ref{eq:LangWABC2}) converges to the initial-value problem for the WABC flow (\ref{WABSeq}). We will refer to this regularisation as 'Cauchy-WABC'. Such stochastic regularisations are often studied in probability theory \cite{gradinaru_singular_2001,attanasio_zero-noise_2009,franco_flandoli_topics_2013}. 

\subsection{Spontaneous stochasticity}

Both regularised initial value problems (\ref{eq:LangWABC}) and (\ref{eq:LangWABC2}) are well-posed. For each of these problems, the solution $\bm x^{(N)}(t)$ is a stochastic process. However, the source of stochasticity vanishes in the limit $N \to \infty$, providing the limiting ideal initial-value problem
\be \label{eq:LangWABCI}
	\frac{d \bm x}{dt} = \bm u_W\left(\bm x\right), \quad
    \bm x (0) = \bm x_0.
\ee
This problem is formally deterministic but generally ill-posed. Our aim is to understand if and how the regularised solutions select a solution of the ideal ill-posed problem (\ref{eq:LangWABCI}) in the vanishing regularisation limit $N \to \infty$.

We say that the stochastically regularised problem defines a spontaneously stochastic solution of the ideal problem if the following two properties are satisfied:

\begin{itemize}
    \item [(i)] There exists a weak limit $x(t) = \lim_{N \to \infty} x^{(N)}(t)$ of the regularised stochastic solution.
    \item [(ii)] The limit is a nontrivial (not a Dirac delta) stochastic process solving the ideal problem (\ref{eq:LangWABCI}).
\end{itemize}

This means that though the limiting equation (\ref{eq:LangWABC}) is deterministic, the solutions are still stochastic in the limit $N \to \infty$. This stochasticity is possible provided that the ideal problem (\ref{eq:LangWABCI}) is ill-posed, and the vanishing noise limit defines a probability distribution on the set its non-unique solutions.

\section{Numerical method}

For the Langevin-WABC regularisation, we use a stochastic method to integrate $N_p$ Lagrangian trajectories following equation (\ref{eq:LangWABC}). For the Cauchy-WABC regularisation (\ref{eq:LangWABC2}) we set the noise part to zero and select a random initial condition for each realization.  
For numerical integration, we use the stochastic solver 'SRA3' from the SciML library, programmed in Julia. This method relies on an optimised Runge-Kutta scheme of strong order $1.5$ \cite{rackauckas_adaptive_2017} and claimed to be of weak order $3$. Integration is performed with a fixed time step $dt_N$ to make sure that it scales properly with cut-off $N$. A series of sampling times $t_i$ is used to compare the different simulations. Since the $t_i$ are not necessarily proportional to $dt_N$, a linear interpolation is performed in order to get the positions at any $t_i$.

On a dimensional argument, it would be natural to set $dt_N \sim \omega_N^{-1}$. However this leads to incorrect results, since the magnitude of displacements $\delta x$ are dominated by large rather than small scales, leading to the so-called sweeping effect. Since the WABC model does have a zero average velocity field, any sweeping effect is local, due to the specific construction by layers. Local sweeping effect in such models has been earlier observed in \cite{thomson_particle_2005, eyink_suppression_2013}, which can lead to incorrect pair-dispersion law as initially found by Elliott and Majda \cite{elliott_pair_1996}. Considering that the correct displacement should be $\delta x \sim k_N^{-1}$, we thus set the time step to $dt_N = a k_N^{-1}$.

We define the ensemble of initial positions for the Cauchy-WABC case as a ball of size $\eta_N$, that we define as
\be
	\eta_N = \frac{10}{k_N}.
\ee
We also define the diffusion term in equation (\ref{eq:LangWABC}) for the Langevin-WABC case as
\be \label{eq:NoiseSca}
\kappa_N  = b^2 \frac{\omega_N^2}{k_N^{2.4}},
\ee
where $b = \sqrt{a}$. Since $dW \sim \frac{1}{\sqrt{a}}$, the noise term explodes when sending $a \to 0$. Thus, $b$ acts as a corrective factor to counterbalance the diverging effect of $a$. This ensures that the noise always dominates at the cut-off scale $k_N^{-1}$.
This particular noise scaling is the result of an optimisation procedure in which we compared probability distributions from simulations of Cauchy-WABC and Langevin-WABC with $h = 1/3$. Further details can be found in Appendix \ref{Ap:Noise}.

\section{Results}

We present here the results of the simulated trajectories starting from ${\bf x}_0 = (3.3492, \, 2.8988, \, 0.7665)$, with $h = 1/3$. From the analysis of weak convergence given in Appendix \ref{Ap:AccMeth}, we set $a = 0.012$ and $N_p = 2^{20} \approx 10^6$. 

\begin{figure}
	\centering
	\begin{subfigure}[t]{0.45\textwidth}
		\centering
		\includegraphics[width=9cm, right]{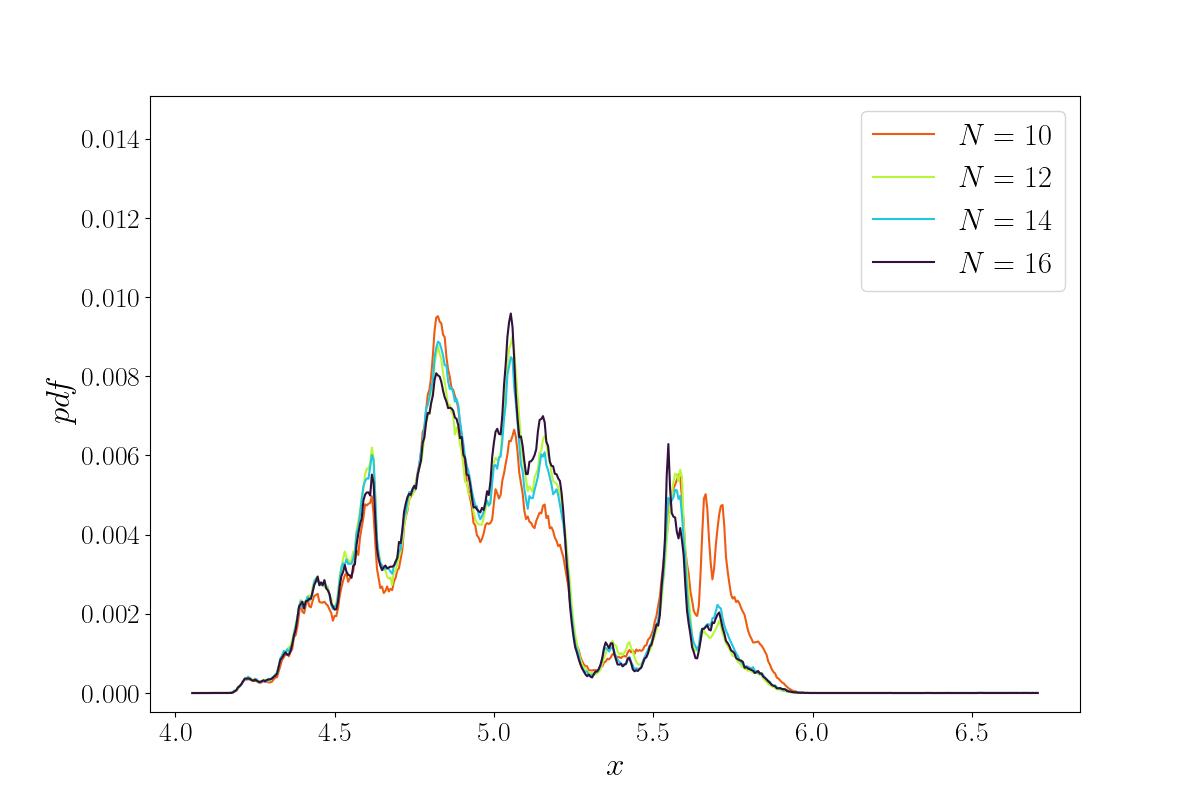}
	\end{subfigure}
	\begin{subfigure}[t]{0.45\textwidth}
		\centering
		\includegraphics[width=9cm, left]{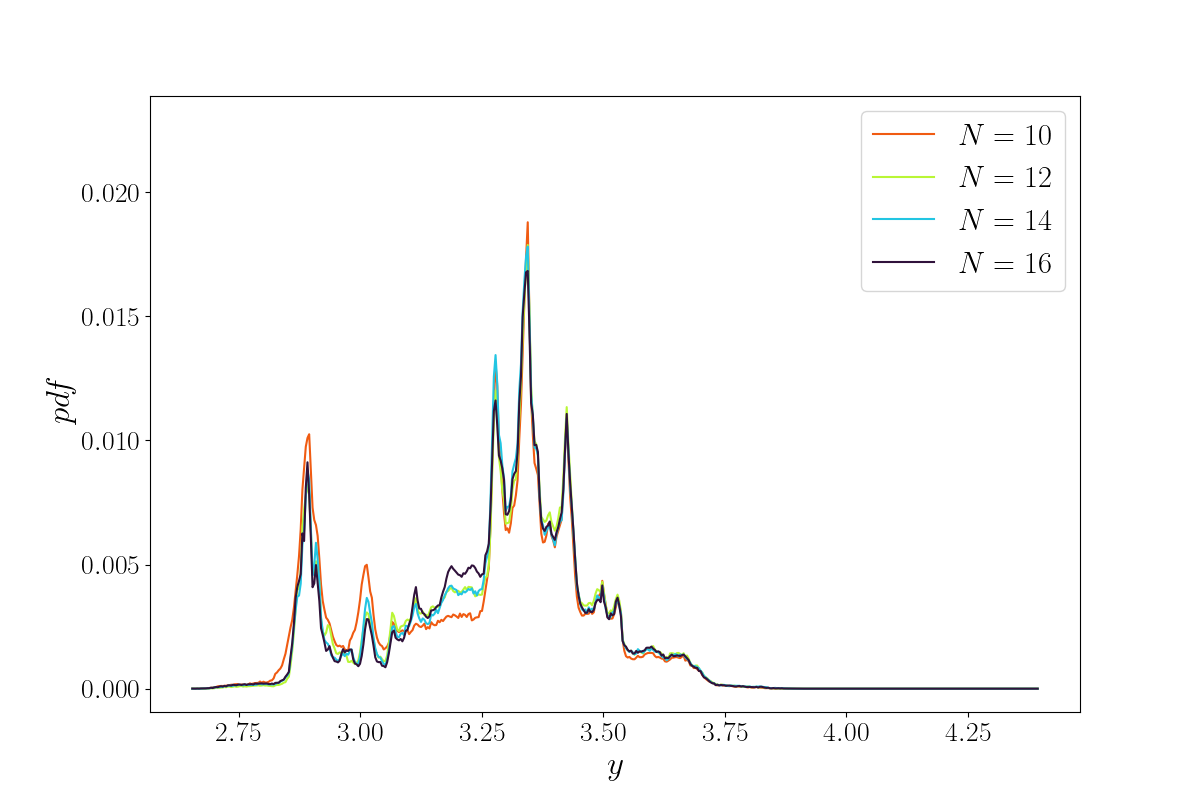}
	\end{subfigure}
	\begin{subfigure}[t]{0.45\textwidth}
		\centering
		\includegraphics[width=9cm, center]{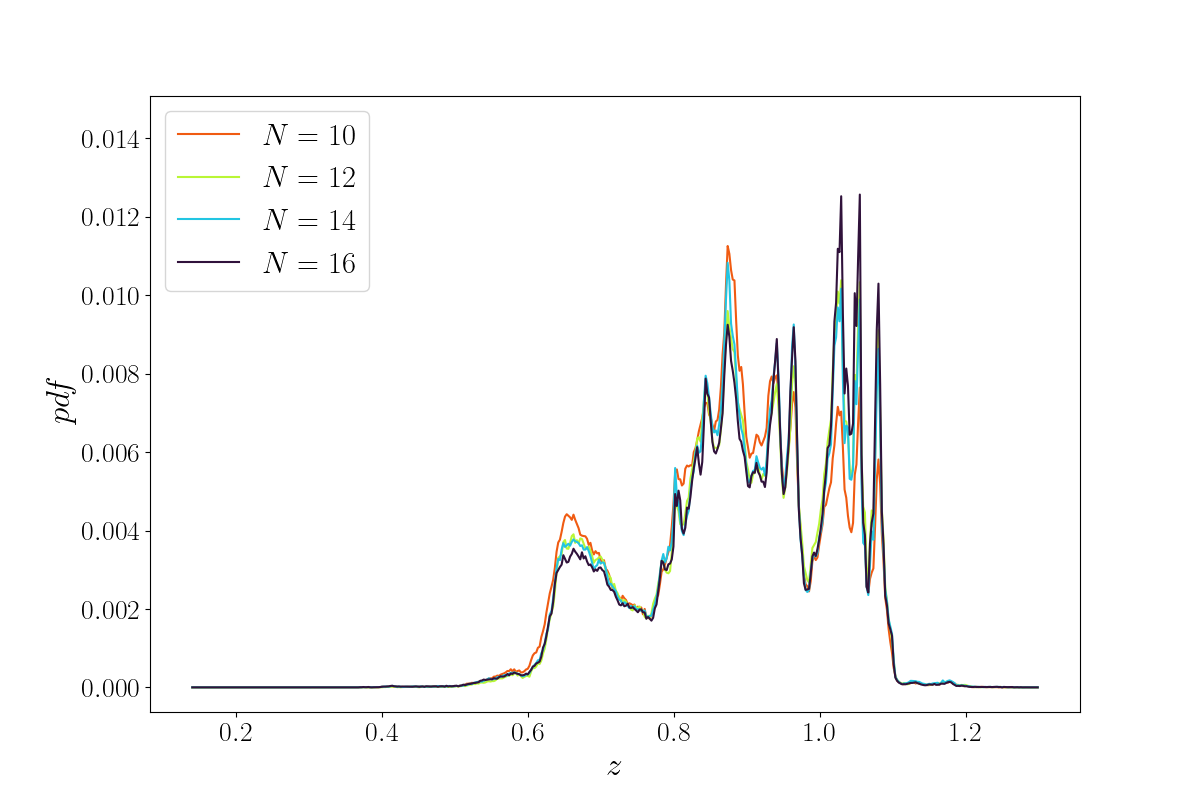}
	\end{subfigure}
	
	\caption{Evolution of one-point statistics through the number of modes $N$ for the Langevin-WABC problem with $h=1/3$ on the three different axes $x$, $y$ and $z$.}
	\label{fig:Hist1DConvMode}
\end{figure}

We test the properties of spontaneous stochasticity by analyzing statistics of particle coordinates at different times and different cutoff parameters $N$. 
We show here the evolution of one and two-points statistics as the number of modes $N$ increases for the Langevin-WABC simulation. Figure \ref{fig:Hist1DConvMode} introduces results of the one-point statistics for the different axes $x$, $y$ and $z$. The $x-y$ slice for the two-points statistics is represented in Figure \ref{fig:Hist2DConvModeXY}. The other two slices are given in Appendix \ref{Ap:AddMat}. Both one-point and two-points statistics demonstrate a numerical convergence for increasing $N$. More specifically, the limiting probability distributions become clearly apparent for $N \geq 12$. These limit distributions appear to be notably non-trivial.

\begin{figure}
	\begin{subfigure}[t]{0.5\textwidth}
		\captionsetup{width=0.90\textwidth,margin={0mm,0.3\textwidth}}
		\centering
		\includegraphics[width=10cm, right]{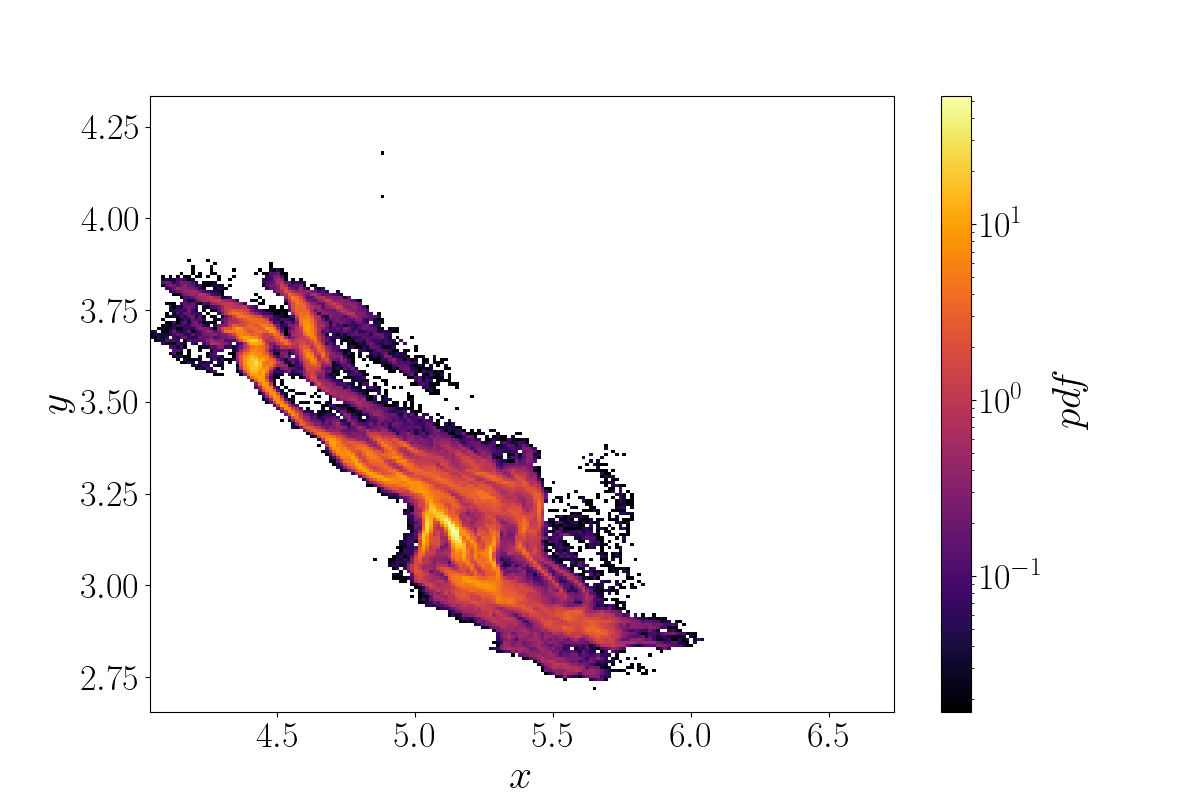}
		\caption{$N = 10$}
	\end{subfigure}
	\begin{subfigure}[t]{0.5\textwidth}
		\captionsetup{width=0.90\textwidth,margin={0.3\textwidth,0mm}}
		\centering
		\includegraphics[width=10cm, left]{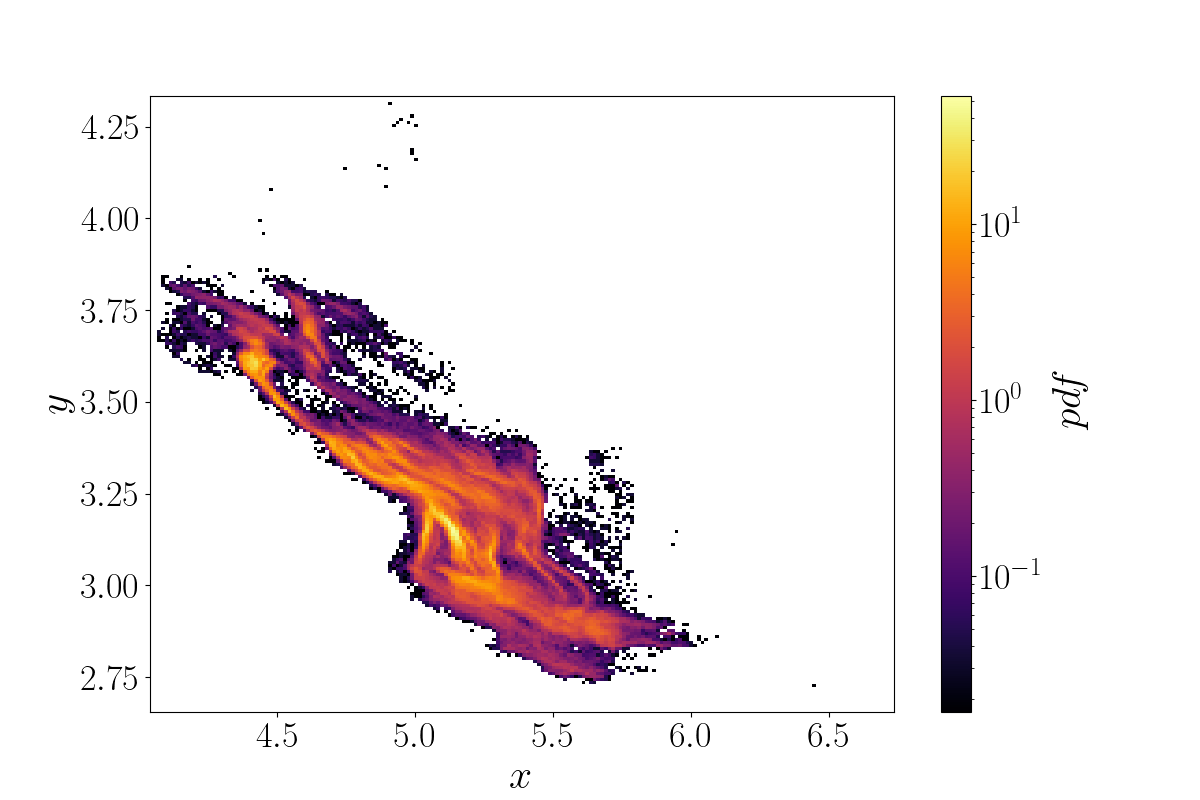}
		\caption{$N = 12$}
	\end{subfigure}
	\begin{subfigure}[t]{0.5\textwidth}
		\captionsetup{width=0.90\textwidth,margin={0mm,0.3\textwidth}}
		\centering
		\includegraphics[width=10cm, right]{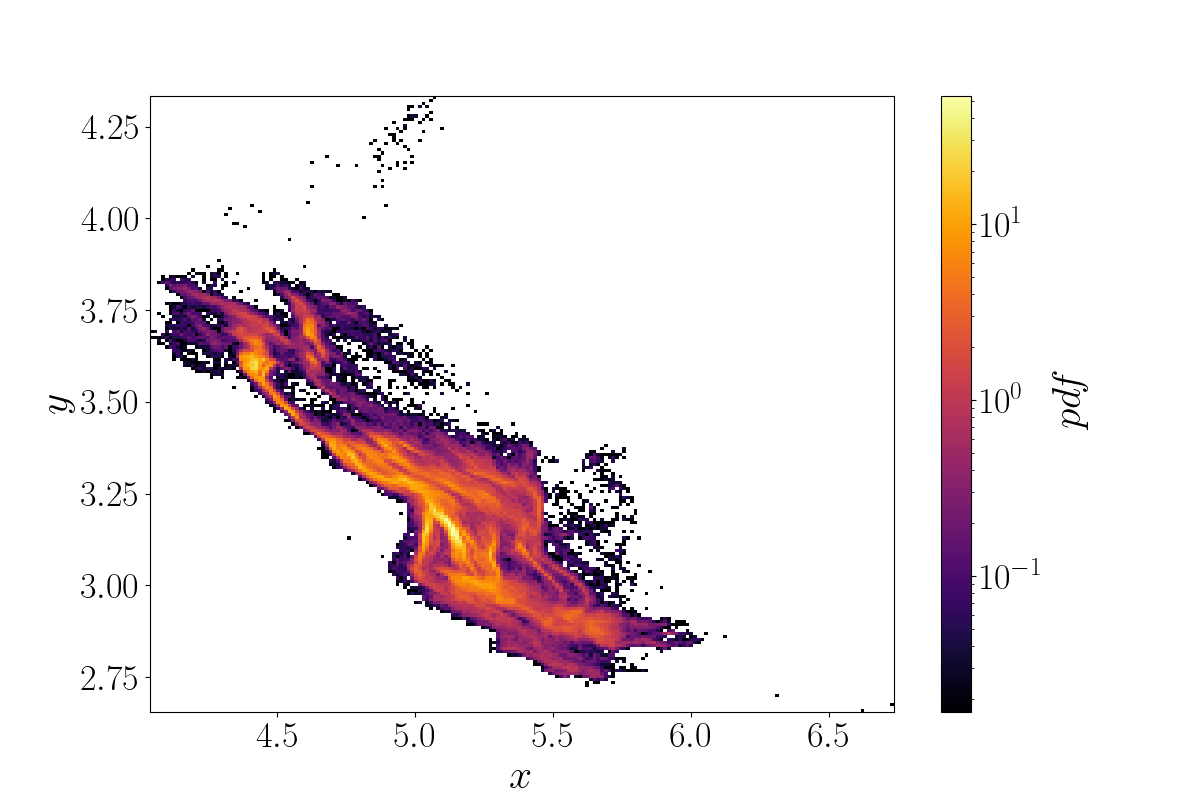}
		\caption{$N = 14$}
	\end{subfigure}
	\begin{subfigure}[t]{0.5\textwidth}
		\captionsetup{width=0.90\textwidth,margin={0.3\textwidth,0mm}}
		\centering
		\includegraphics[width=10cm, left]{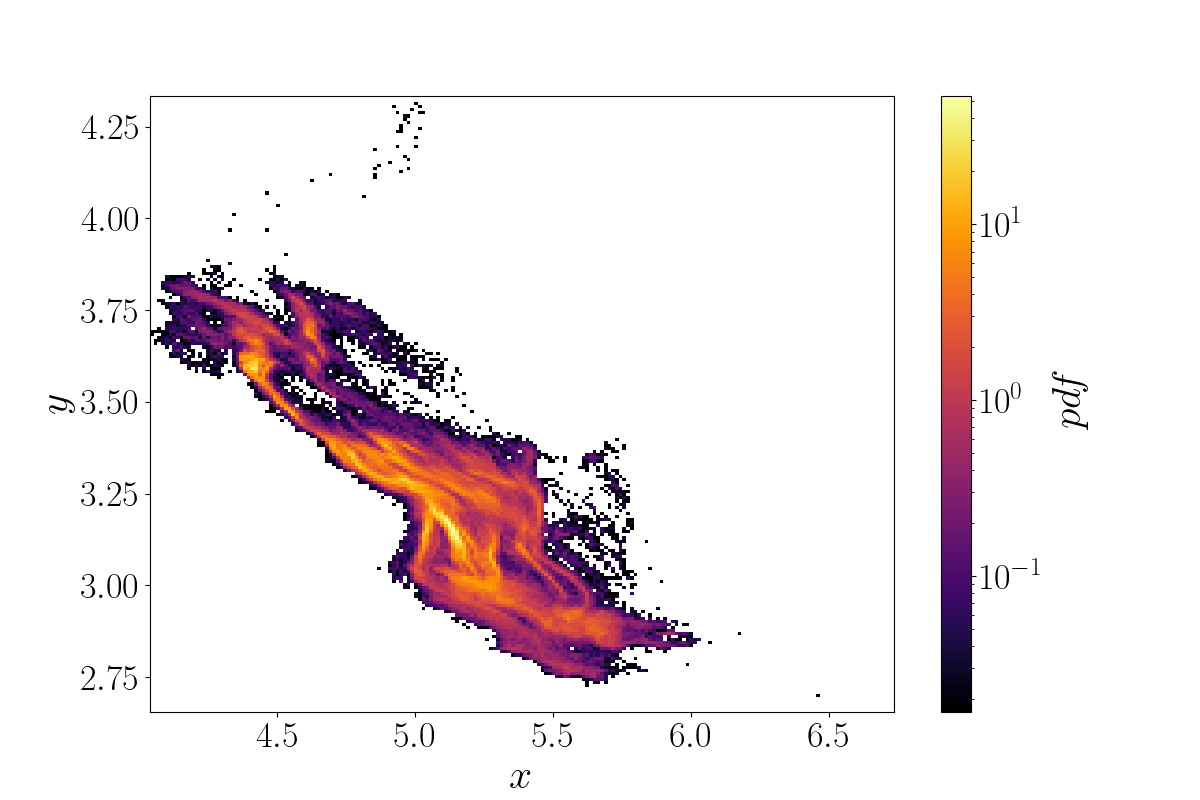}
		\caption{$N = 16$}
	\end{subfigure}
	\caption{Evolution of two-points statistics through the number of modes $N$ for the Langevin-WABC problem for the XY-slice and $h = 1/3$.}
	\label{fig:Hist2DConvModeXY}
\end{figure}

\begin{figure}
	\centering
	\includegraphics[width=12cm]{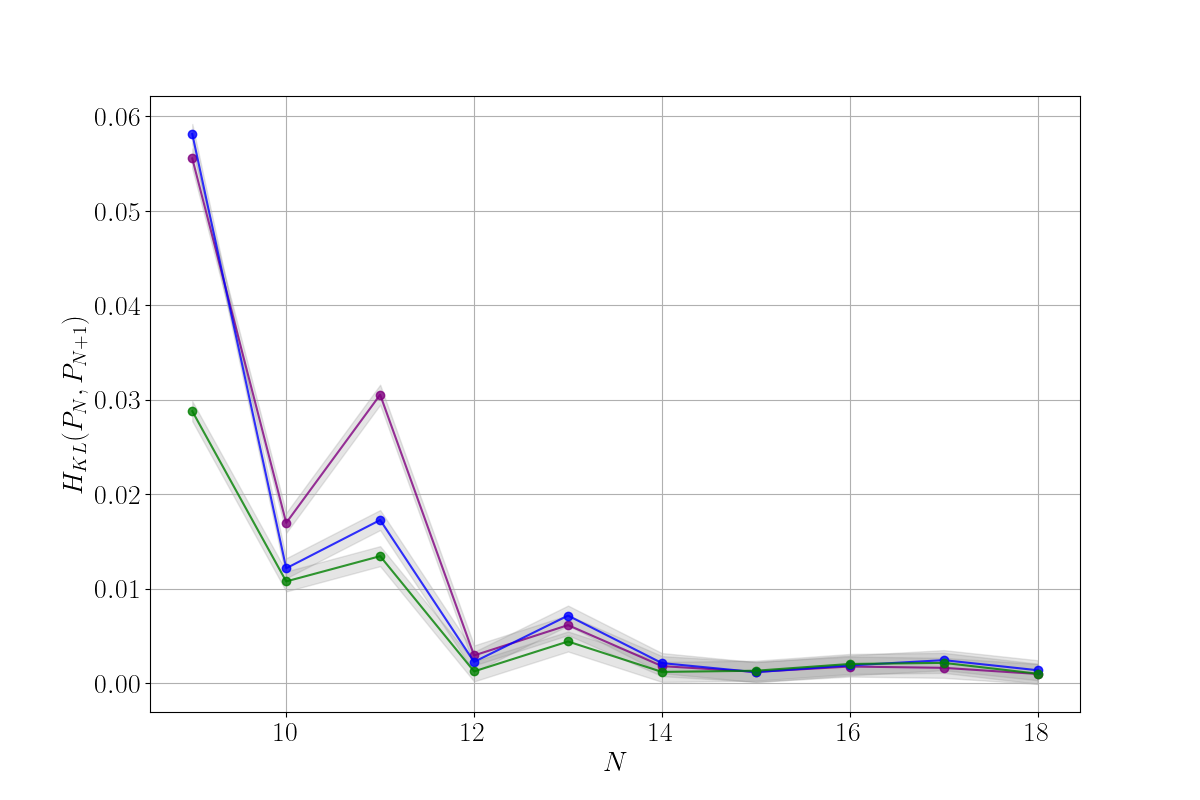}
	
	\caption{Evolution of the Kullback-Leibler divergence between one-point statistics of increasing number of modes $N$ as a function of $N$ for Langevin-WABC with $h=1/3$ for the $x$-axis (purple), the $y$-axis (blue) and the z-axis (green).}
	\label{fig:HKLConv}
\end{figure}

We quantify the convergence of the one-point statistics using the Kullback-Leibler divergence $H_{KL}$. Being a divergence, we recall that for two probability distributions $p$ and $q$, we have $H_{KL}(p, q) = 0$ if and only if $p = q$.
The results regarding the convergence of the one-point statistics are presented in Figure \ref{fig:HKLConv} for axes $x$, $y$ and $z$. All curves decrease as $N$ increases, and reach a plateau for $N \geq 14$. As discussed in Appendix \ref{Ap:AccMeth}, this saturation corresponds to finite-size effects, and are mainly dues to our finite number of particles.

\begin{figure}
	\centering
	\includegraphics[width=12cm]{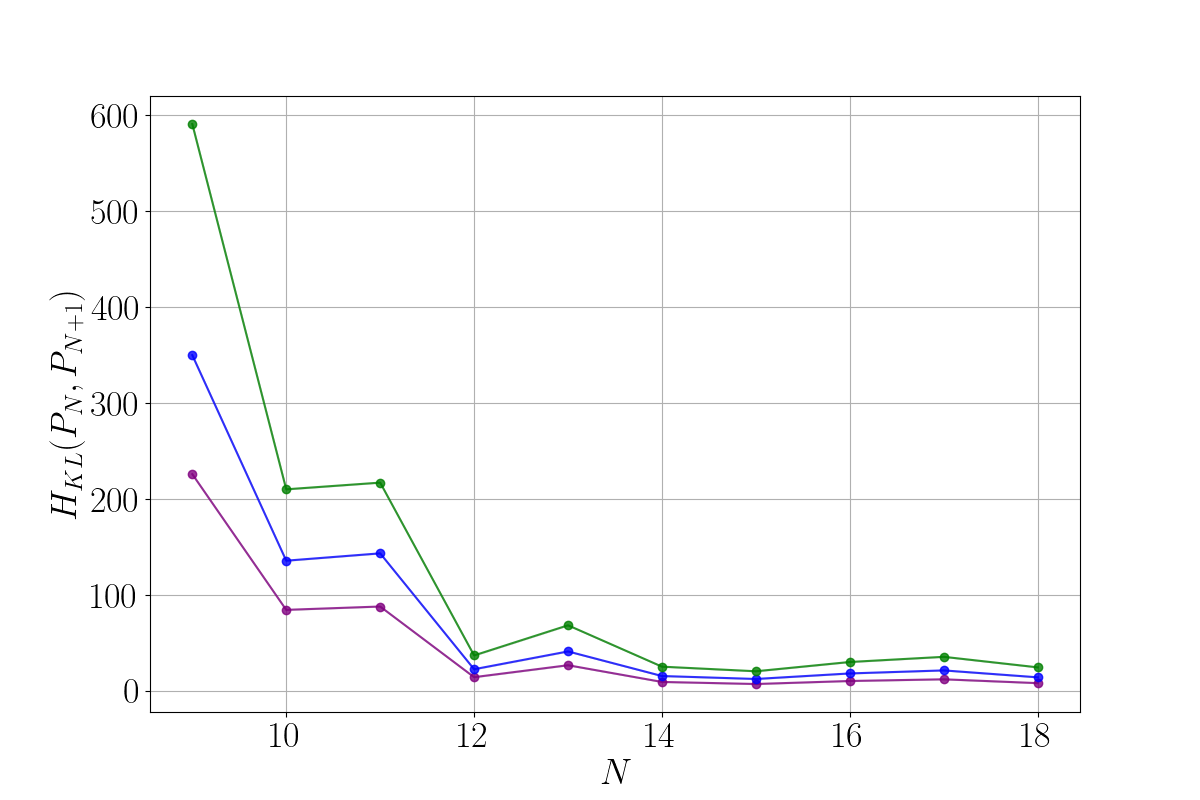}
	
	\caption{Evolution of the Kullback-Leibler divergence between two-points statistics of increasing number of modes $N$ as a function of $N$ for Langevin-WABC with $h=1/3$ for the $x-y$-slicce (purple), the $y-z$-slice (blue) and the $x-z$-slice (green).}
	\label{fig:HKLConv2D}
\end{figure}

The convergence of the two-points statistics can also be quantified using the Kullback-Leibler divergence. The corresponding results are depicted in Figure \ref{fig:HKLConv2D}. The behaviour is qualitatively similar to what is obtained for the one-point statistics, with all curves decreasing as $N$ increases until a plateau is reached. However, due to the curse of dimensionality, the finite size errors are larger and induce larger values of the plateau in the  $2d$ case.

\section{Discussion}

\subsection{Conclusions}
The results, obtained for the Langevin-WABC simulation with $h = 1/3$, show a convergence of statistics in the limit of numerical errors. The limit one-point and two-points distributions are highly non-trivial and distinct from simple Diracs. From those results, we conclude that the WABC model can build spontaneous stochasticity in the sense that Lagrangian trajectories appear to be still stochastic in the limit of vanishing noise.

Computational burdens limits us to $N_p = 2^{20}$, which is, at the present, the main limiting factor to the convergence of the Kullback-Leibler divergence. In the event where  higher values of $N_p$ could be reached, the scheme accuracy would become the main limiting factor. We finally note that the curse of dimensionality also limits the accuracy of the two-points statistics. A larger number of particles could solve this issue but is in practice too difficult to perform.

\subsection{Influence of regularisation}
\begin{figure}
	\centering
	\begin{subfigure}[t]{0.45\textwidth}
		\centering
		\includegraphics[width=9cm, right]{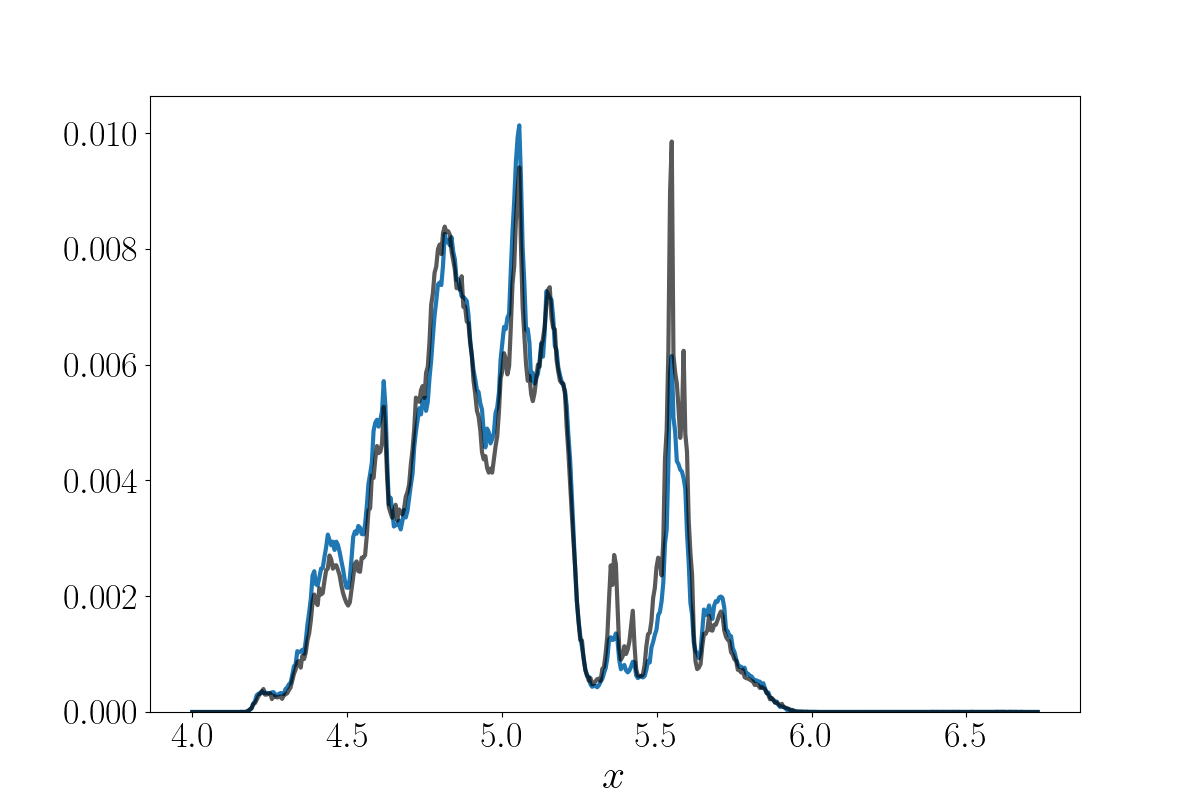}
	\end{subfigure}
	\begin{subfigure}[t]{0.45\textwidth}
		\centering
		\includegraphics[width=9cm, left]{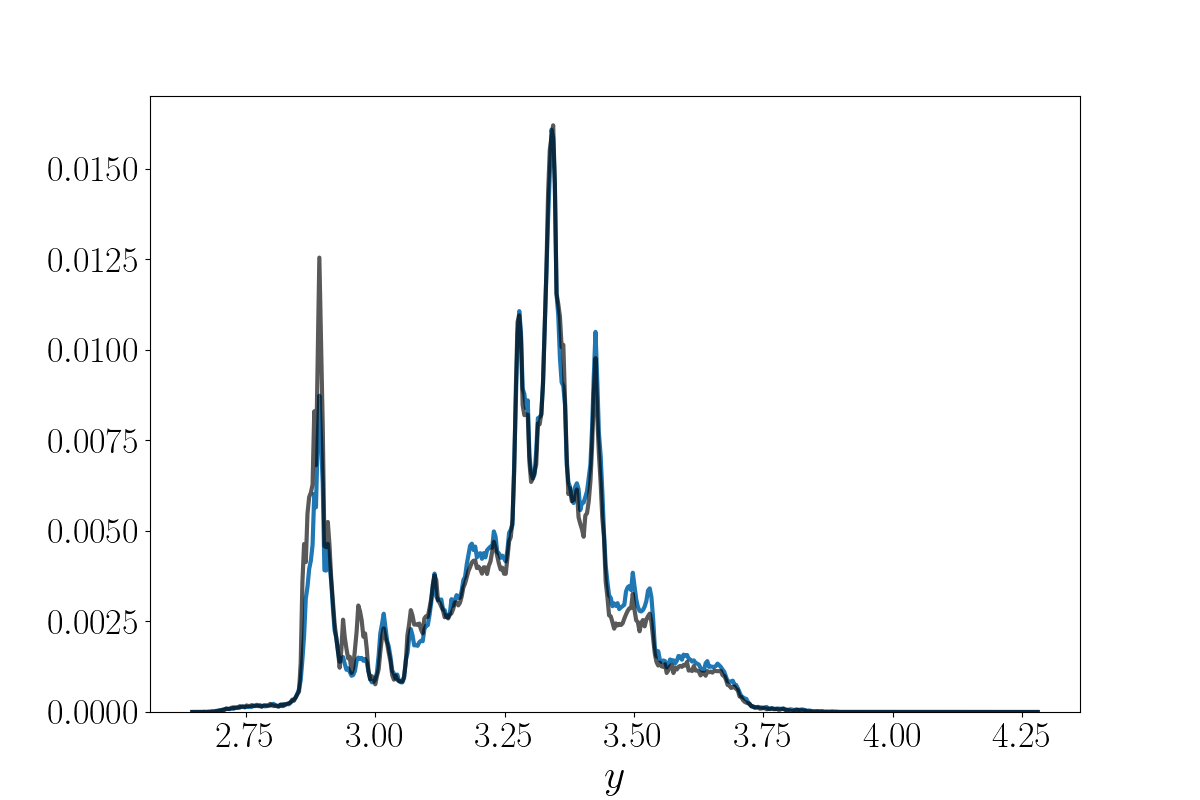}
	\end{subfigure}
	
	\begin{subfigure}[t]{0.45\textwidth}
		\centering
		\includegraphics[width=9cm, center]{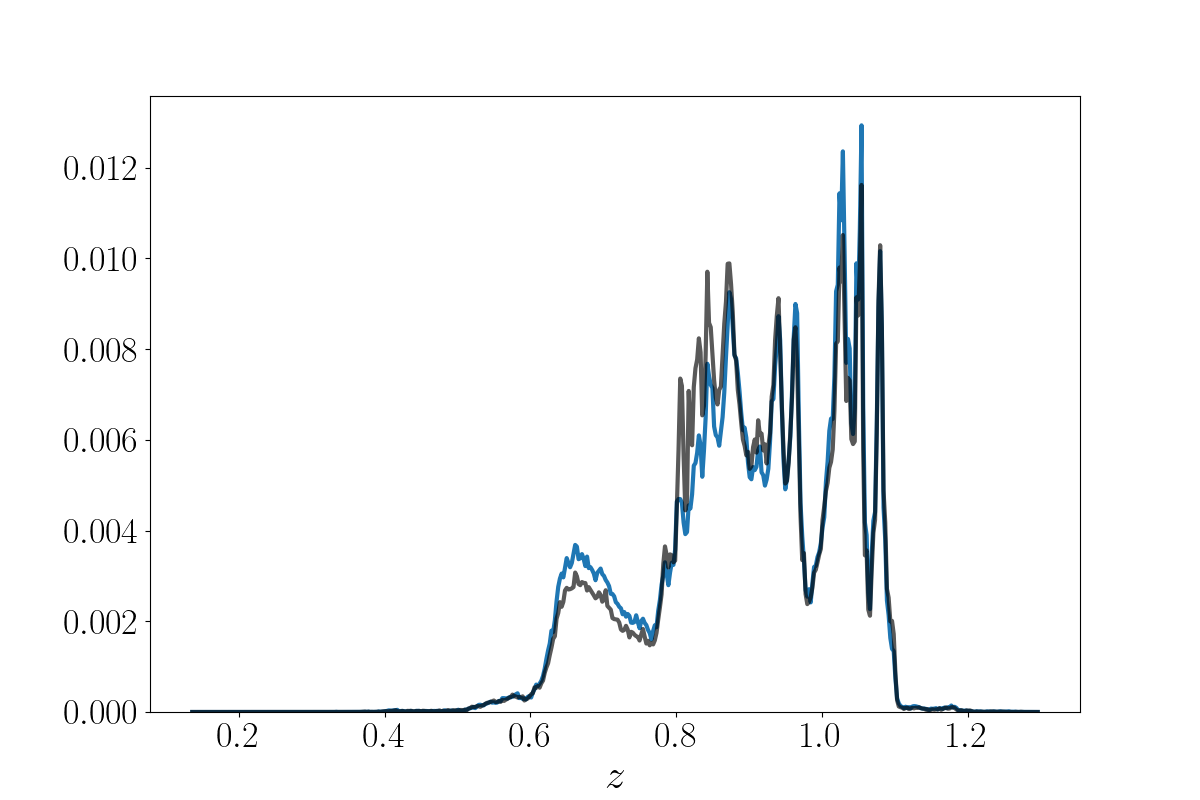}
	\end{subfigure}
	
	\caption{Comparison of one point statistics between Cauchy-WABC (black) and Langevin-WABC (blue) with $h=1/3$ and $N=16$ for the $x$-axis (top left), the $y$-axis (top right) and the $z$-axis (bottom).}
	\label{fig:NoiseComph3}
\end{figure}

\begin{figure}
	\centering
	\includegraphics[width=12cm]{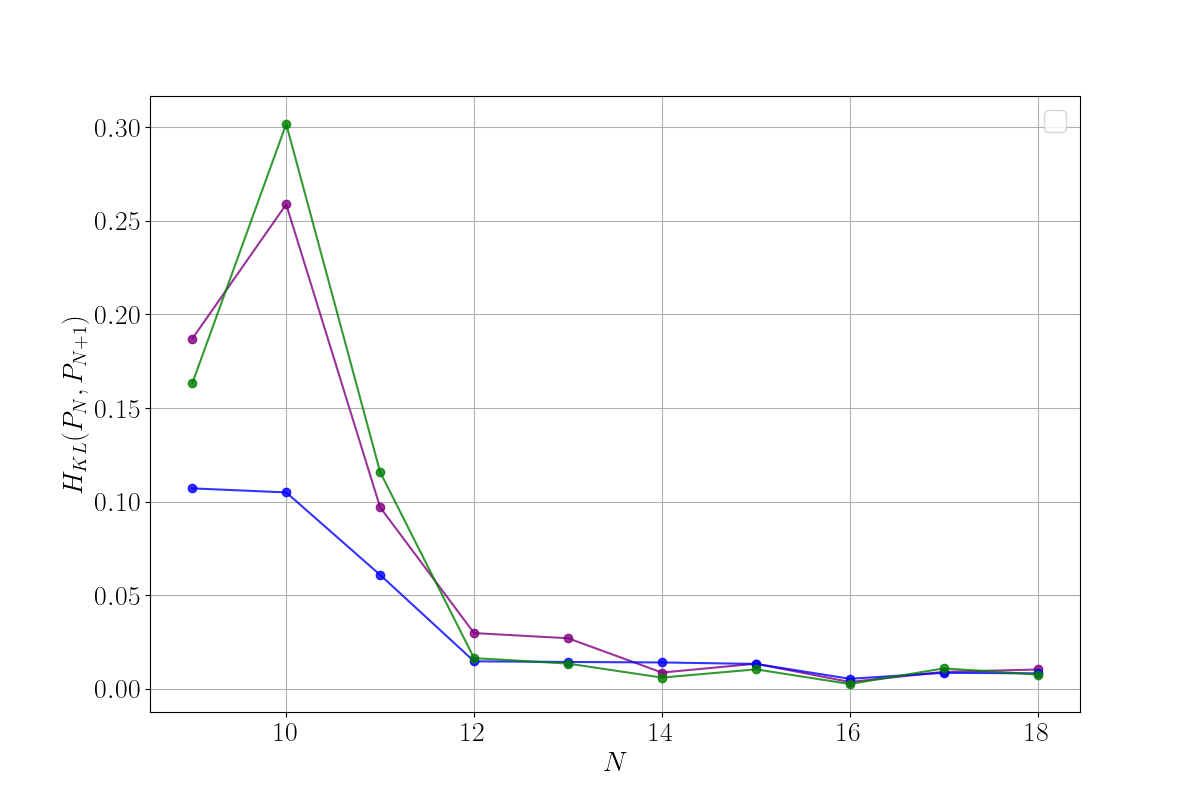}
	
	\caption{Evolution of the Kullback-Leibler divergence between one-point statistics of increasing number of modes $N$ as a function of $N$ for Cauchy-WABC with $h=1/3$ for the $x$-axis (purple), the $y$-axis (blue) and the z-axis (green).}
	\label{fig:HKLConvCauchy}
\end{figure}

Switching regularisation to the Cauchy-WABC does not change the results. We indeed show in Figure \ref{fig:NoiseComph3} a comparison between the Cauchy-WABC and the Langevin-WABC one-point distributions. We also show in Figure \ref{fig:HKLConvCauchy} the Kullback-Leibler divergence of the one-point statistics as $N$ increases.

As in the Langevin-WABC case, we observe that the curves decrease until reaching a plateau, showing the same convergence through the number of modes. The limit probability distributions appear to be identical between the Cauchy-WABC and the Langevin-WABC cases, in compliance with what was observed in the Navier--Stokes simulations~\cite{thalabard_butterfly_2020} and other turbulence models~\cite{mailybaev2016spontaneously,mailybaev_spontaneously_2023,bandak_spontaneous_2024}. No other regularisation was tested, but we expect that the same spontaneously stochastic limit holds for a large class of regularisations.

In this paper we neither review the influence of the Hölder exponent nor the initial position on the observation of spontaneous stochasticity. We expect it to appear for any $h < 1$ and for any initial position. The numerical investigation of such claim is however tedious,  particularly for $h$ closer to $1$. It indeed appears that the time at which spontaneous stochasticity appears increases with $h$, which lengthen the simulations. This can be problematic as one would need to decrease the integration precision in order to limit simulation time, possibly altering the results' precision.

\subsection{Influence of chaos?}
It was shown that the ABC model could lead to Lagrangian chaos for well-chosen $A$, $B$ and $C$ parameters. We checked qualitatively that the chosen parameters here correctly display chaos. By extension, one may expect that the  WABC model can also display Lagrangian chaos. 

It is up to now still unclear whether chaos is important to spontaneous stochasticity or not. This model offers the possibility to check this influence. We note that this model can be used to particularly observe the influence of diverging positive Lyapunov exponents onto the emergence of a Richardson-like regime. The evaluation of those exponents can be difficult in the ABC model since it can have multiple attractors. A simpler approach for such investigations could use the instantaneous Lyapunov exponents as indicators of strong dispersive areas. Since this flow is stationary, the computation of those coefficient is simple when using a proper approximation as the one introduced by Nolan et al. \cite{nolan_finite-time_2020}. 

We finally note that such model could be used to investigate transition towards spontaneous stochasticity. The $A$, $B$ and $C$ parameters could be used to trigger or not chaos in the model, eventually leading to phase transitions if such link with spontaneous stochasticity exists.

\section*{Acknowledgements}
We would like to thank G. Eyink for his comments on the numerical implementation and the underlying sweeping effect. We also thank A. Considera and S. Thalabard for all the useful discussions we had.

\begin{appendices}

\section{Additional material} \label{Ap:AddMat}
We present in this appendix the two other directions for the two-points probability distributions in the Langevin-WABC case.

\begin{figure}
	\begin{subfigure}[t]{0.5\textwidth}
		\captionsetup{width=0.90\textwidth,margin={0mm,0.3\textwidth}}
		\centering
		\includegraphics[width=10cm, right]{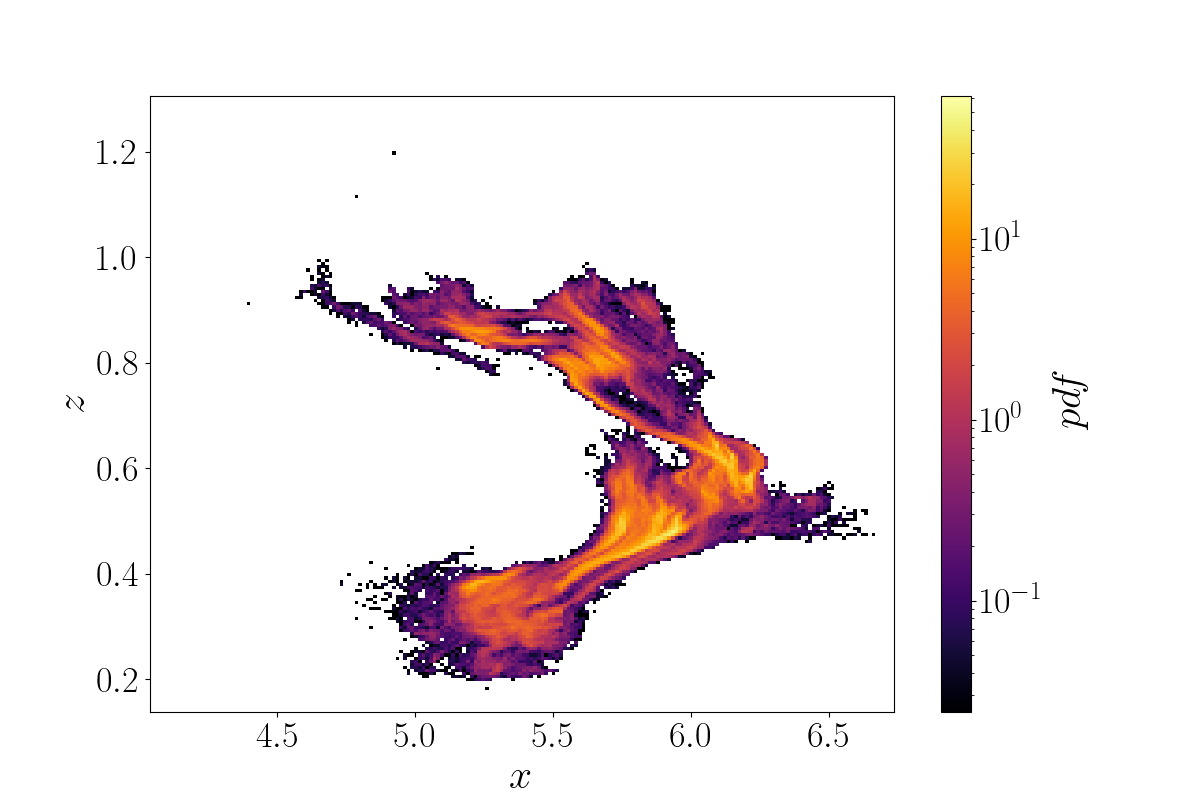}
		\caption{$N = 10$}
	\end{subfigure}
	\begin{subfigure}[t]{0.5\textwidth}
		\captionsetup{width=0.90\textwidth,margin={0.3\textwidth,0mm}}
		\centering
		\includegraphics[width=10cm, left]{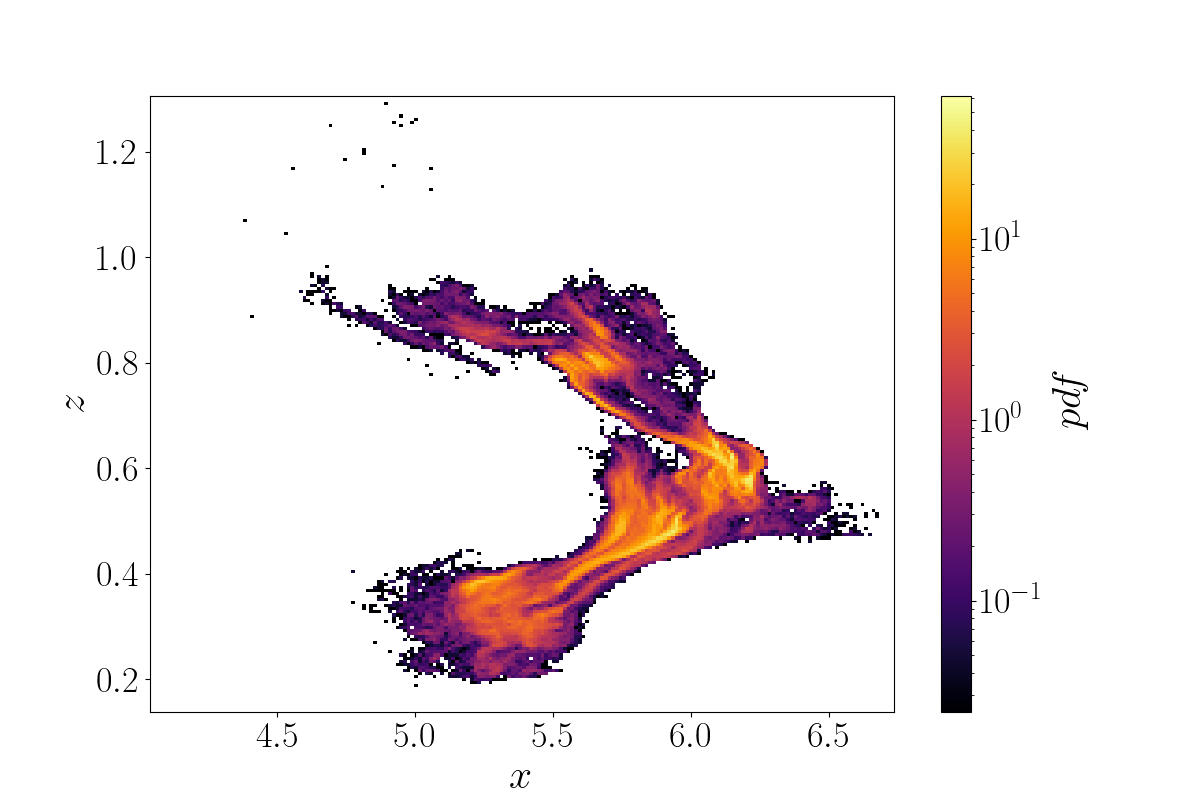}
		\caption{$N = 12$}
	\end{subfigure}
	\begin{subfigure}[t]{0.5\textwidth}
		\captionsetup{width=0.90\textwidth,margin={0mm,0.3\textwidth}}
		\centering
		\includegraphics[width=10cm, right]{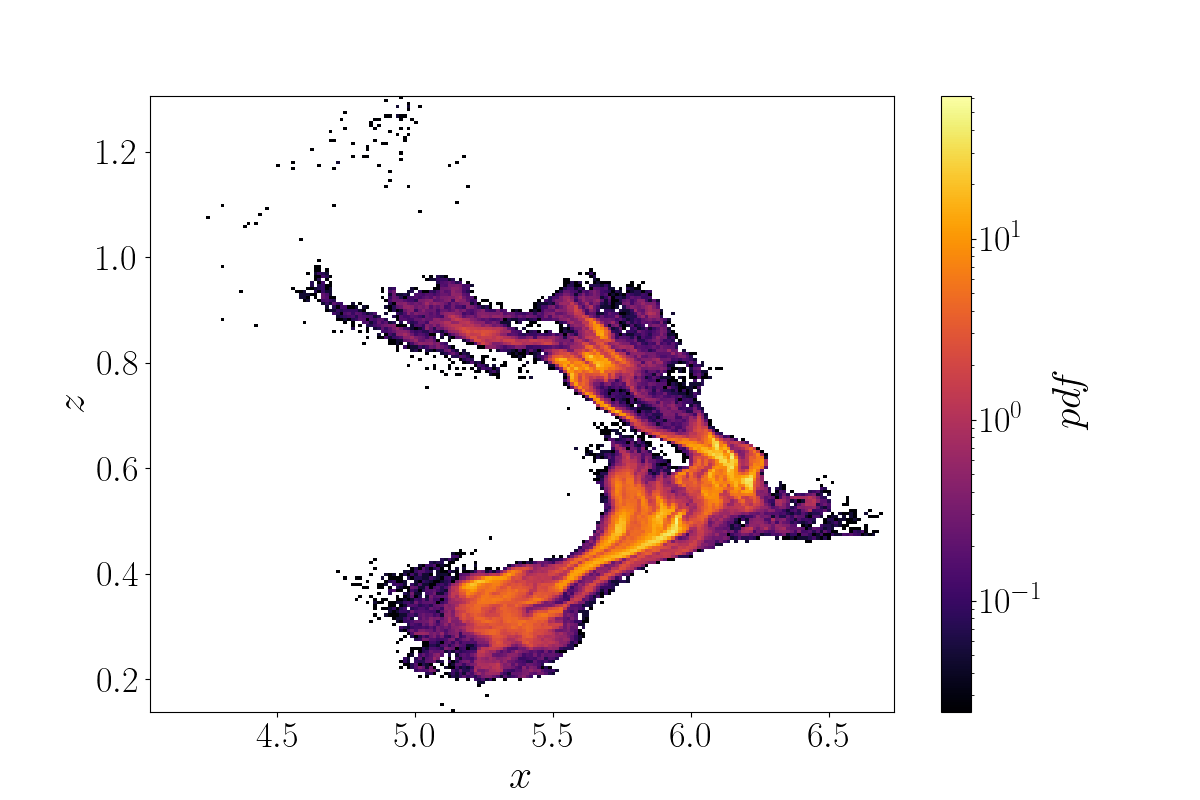}
		\caption{$N = 14$}
	\end{subfigure}
	\begin{subfigure}[t]{0.5\textwidth}
		\captionsetup{width=0.90\textwidth,margin={0.3\textwidth,0mm}}
		\centering
		\includegraphics[width=10cm, left]{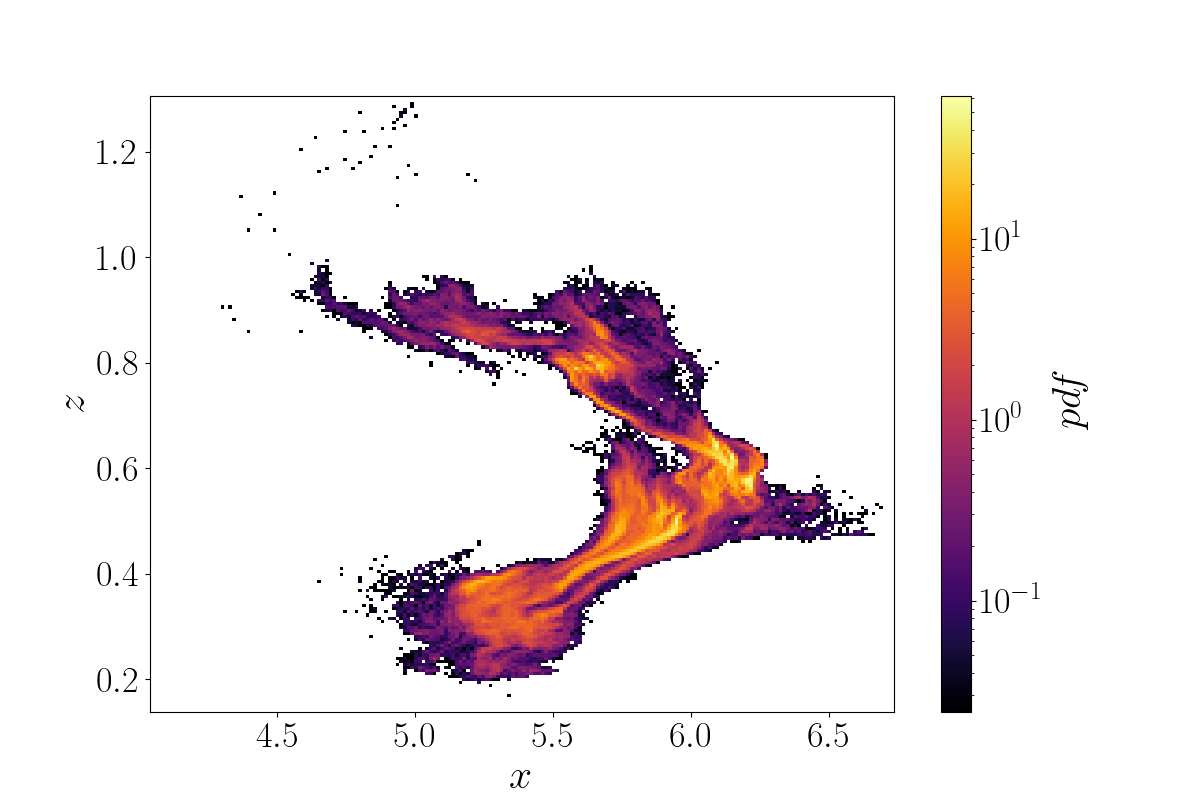}
		\caption{$N = 16$}
	\end{subfigure}
	\caption{Evolution of two-points statistics through the number of modes $N$ for the Langevin-WABC problem for the XZ-slice and $h = 1/3$.}
	\label{fig:Hist2DConvModeXZ}
\end{figure}

\begin{figure}
	\begin{subfigure}[t]{0.5\textwidth}
		\captionsetup{width=0.90\textwidth,margin={0mm,0.3\textwidth}}
		\centering
		\includegraphics[width=10cm, right]{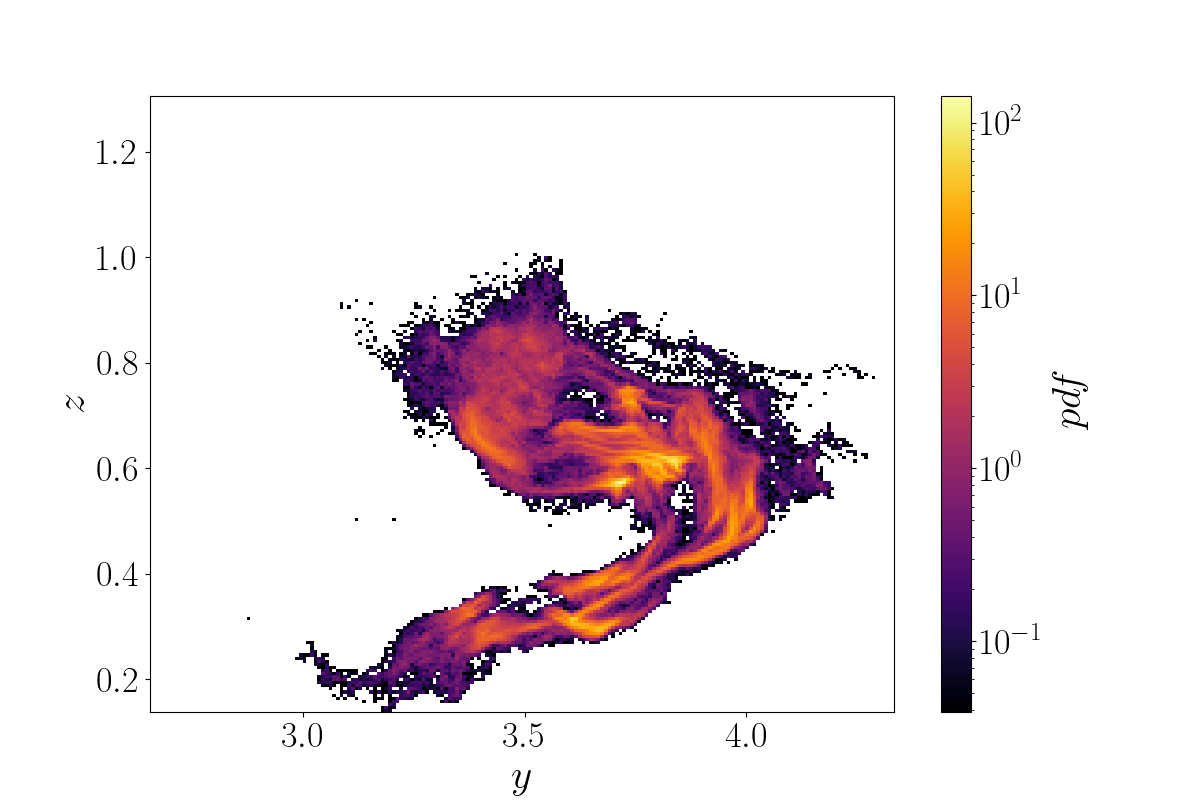}
		\caption{$N = 10$}
	\end{subfigure}
	\begin{subfigure}[t]{0.5\textwidth}
		\captionsetup{width=0.90\textwidth,margin={0.3\textwidth,0mm}}
		\centering
		\includegraphics[width=10cm, left]{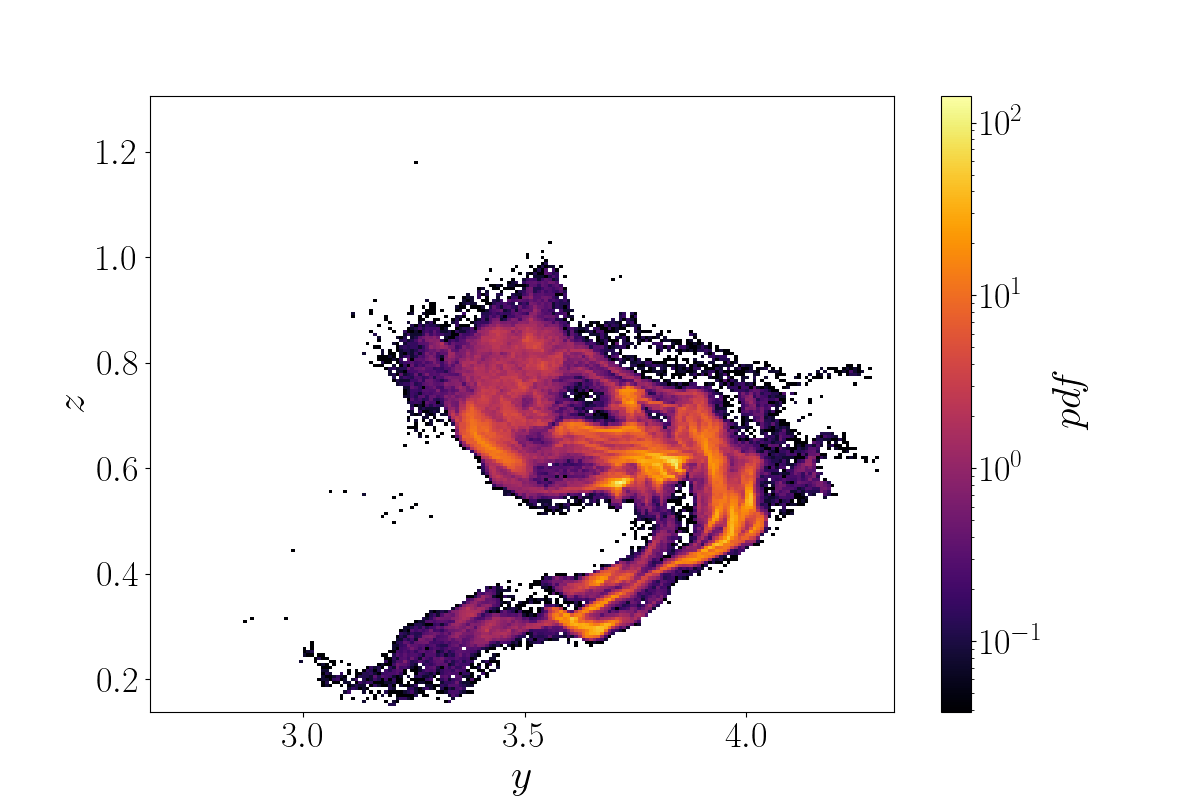}
		\caption{$N = 12$}
	\end{subfigure}
	\begin{subfigure}[t]{0.5\textwidth}
		\captionsetup{width=0.90\textwidth,margin={0mm,0.3\textwidth}}
		\centering
		\includegraphics[width=10cm, right]{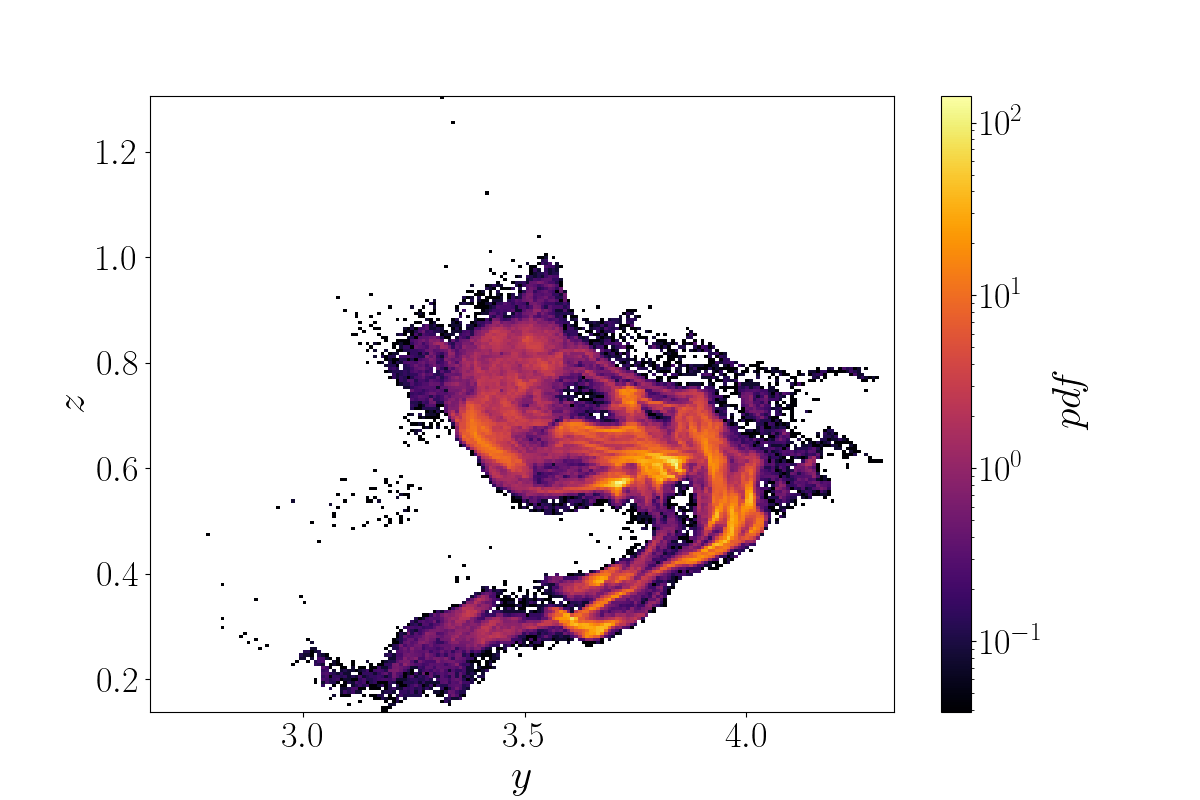}
		\caption{$N = 14$}
	\end{subfigure}
	\begin{subfigure}[t]{0.5\textwidth}
		\captionsetup{width=0.90\textwidth,margin={0.3\textwidth,0mm}}
		\centering
		\includegraphics[width=10cm, left]{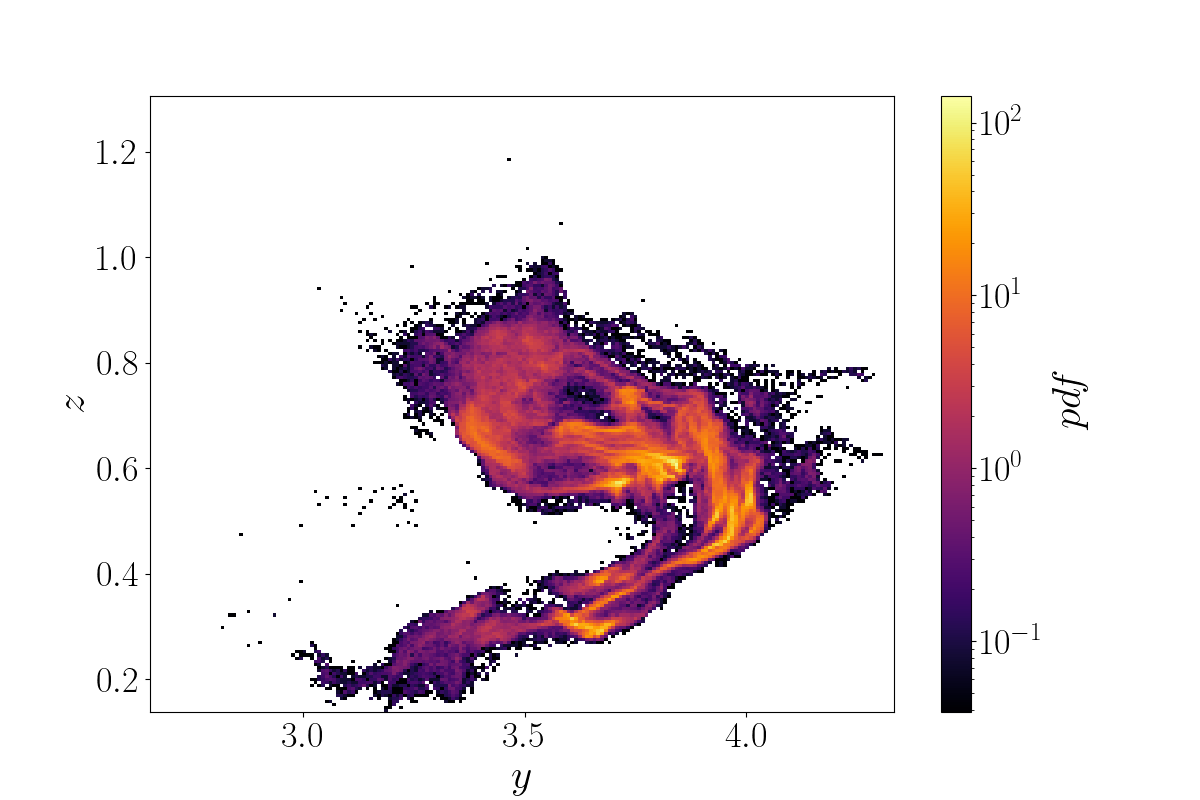}
		\caption{$N = 16$}
	\end{subfigure}
	\caption{Evolution of two-points statistics through the number of modes $N$ for the Langevin-WABC problem for the XY-slice and $h = 1/3$.}
	\label{fig:Hist2DConvModeYZ}
\end{figure}

\section{Accuracy of the methodology} \label{Ap:AccMeth}
We introduce here the accuracy tests of our numerical methodology. We identify two sources that can alter the results: the numerical integration associated with the time time step $a$ and the statistical convergence associated with the number of particles $N_p$. This section is dedicated to quantifying the influence of both factors in our results.

\subsection{Integration precision}
\begin{figure}
	\centering
	\begin{subfigure}[t]{0.8\textwidth}
		\centering
		\includegraphics[width=11cm]{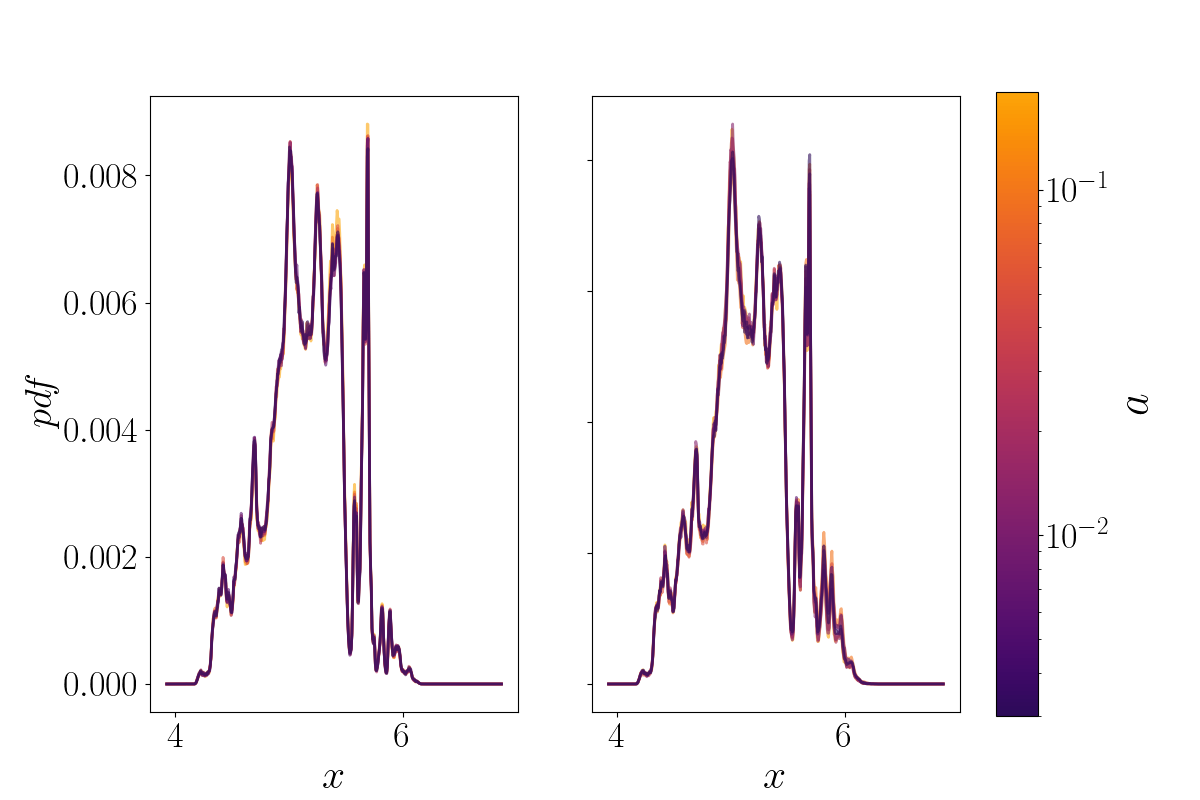}
	\end{subfigure}
	
	\begin{subfigure}[t]{0.8\textwidth}
		\centering
		\includegraphics[width=11cm]{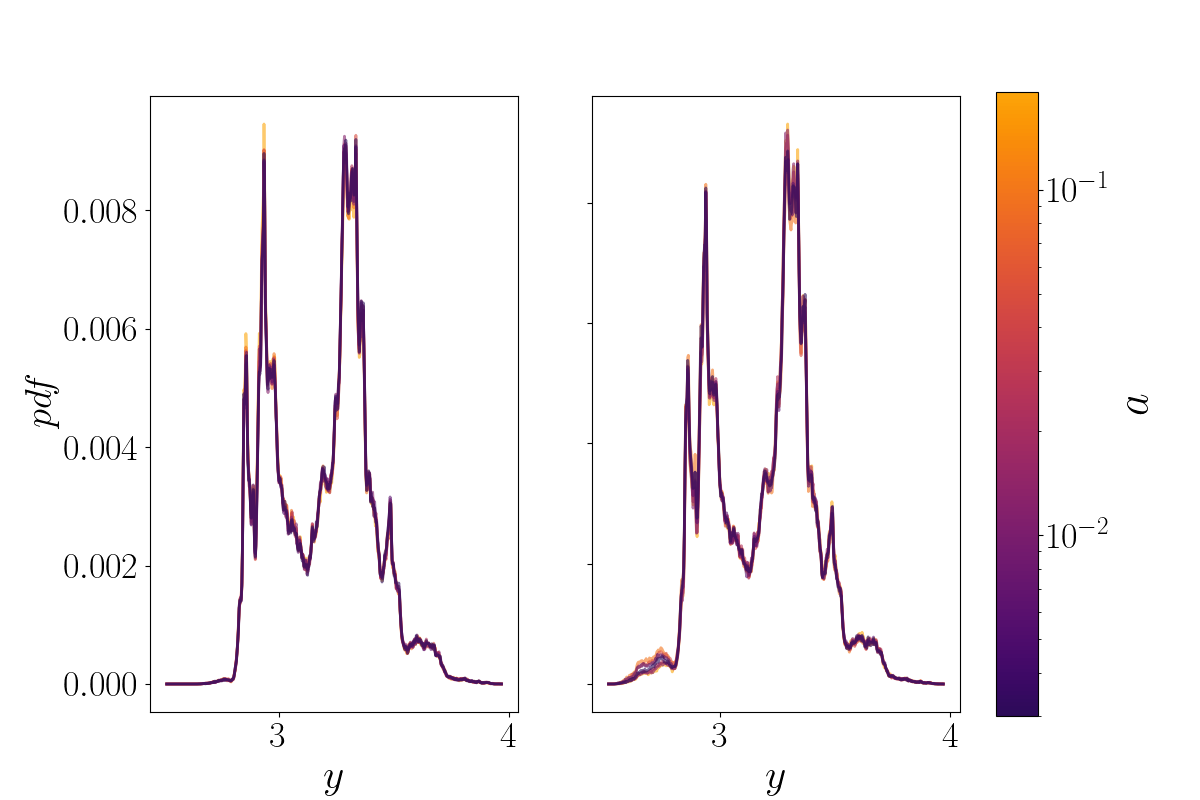}
	\end{subfigure}
	
	\begin{subfigure}[t]{0.8\textwidth}
		\centering
		\includegraphics[width=11cm]{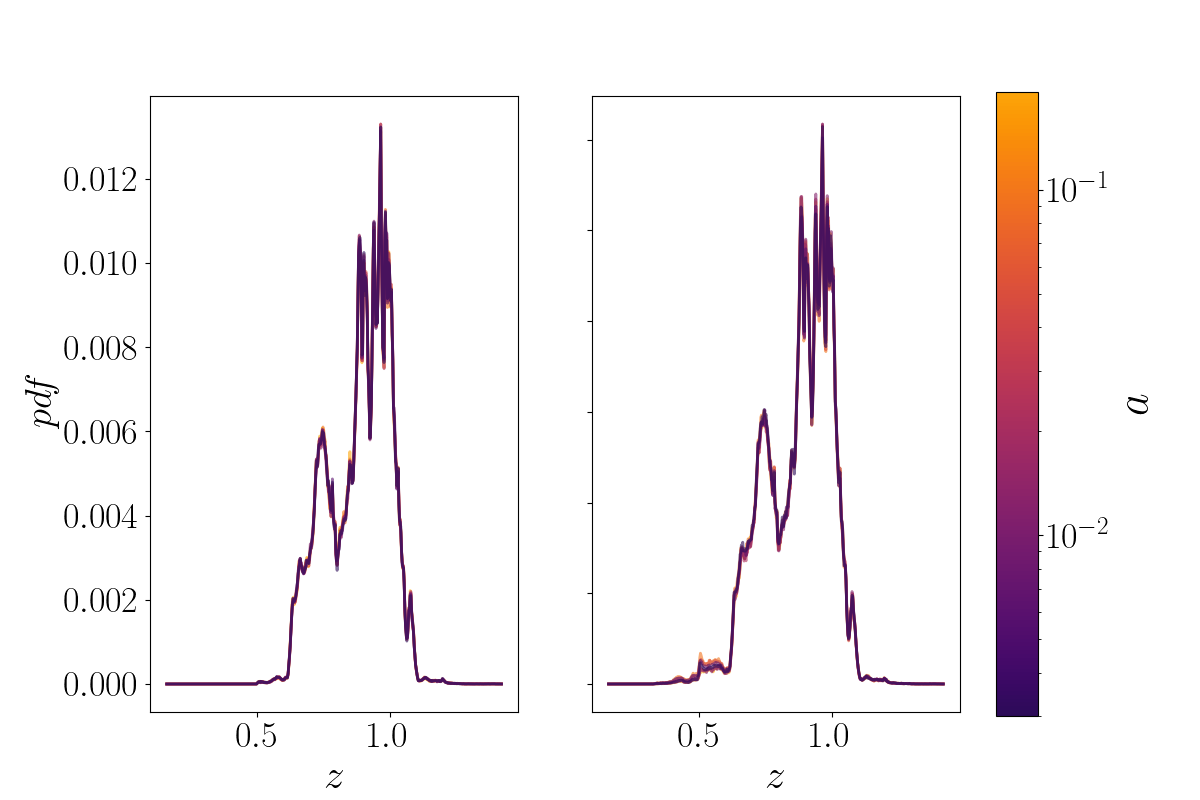}
	\end{subfigure}
	
	\caption{One-point statistics of the Langevin-WABC model where $h=1/3$ and $N = 14$, with decreasing $a$ coefficient: Euler-Mayurama (Left) and SRA3 (Right) algorithms when $b = 1$ and varying $a$ (indicated by the colour bar). We display, from top to bottom, the one-point statistics for the 3 coordinates $x$, $y$ and $z$.}
	\label{fig:AccEMSRA_b1}
\end{figure}

In this scenario, we set the number of particles $N_p = 2^{20}$, the Hölder exponent to $h = 1/3$ and we vary $a$ for the Langevin-WABC setup, changing the time stepping $dt_N = \frac{a}{k_N}$. The corresponding one-point probability distributions for all the simulated $a$ are displayed in Figure \ref{fig:AccEMSRA_b1}. A comparison with the classical Euler-Mayurama algorithm of weak order $1$ is also presented.

We observe that the probability distributions do not seem to change for the range of tested time steps. Changing the solver does not seem to have an impact either on the resulting statistics. This therefore indicates a low influence of the integration precision onto the final statistics.

\subsection{Statistical convergence}
\begin{figure}
	\centering
	\begin{subfigure}[t]{0.45\textwidth}
		\centering
		\includegraphics[width=9cm, right]{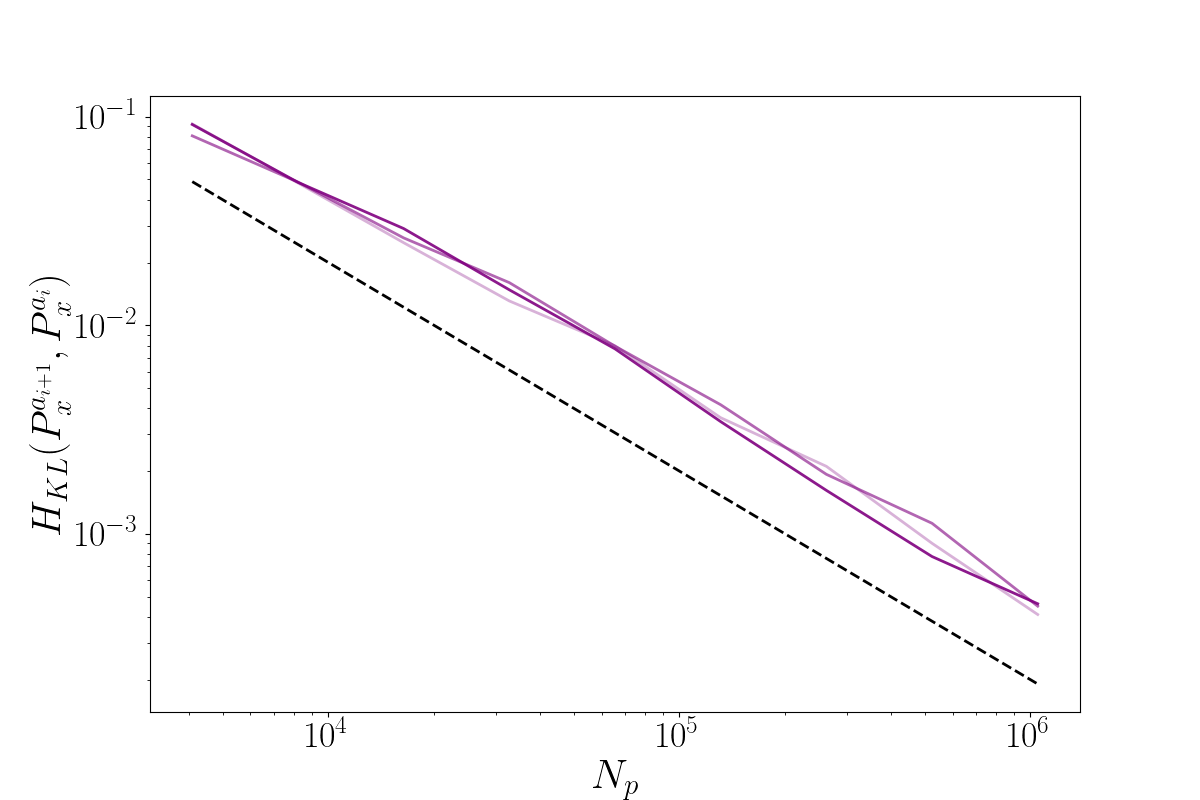}
	\end{subfigure}
	\begin{subfigure}[t]{0.45\textwidth}
		\centering
		\includegraphics[width=9cm, left]{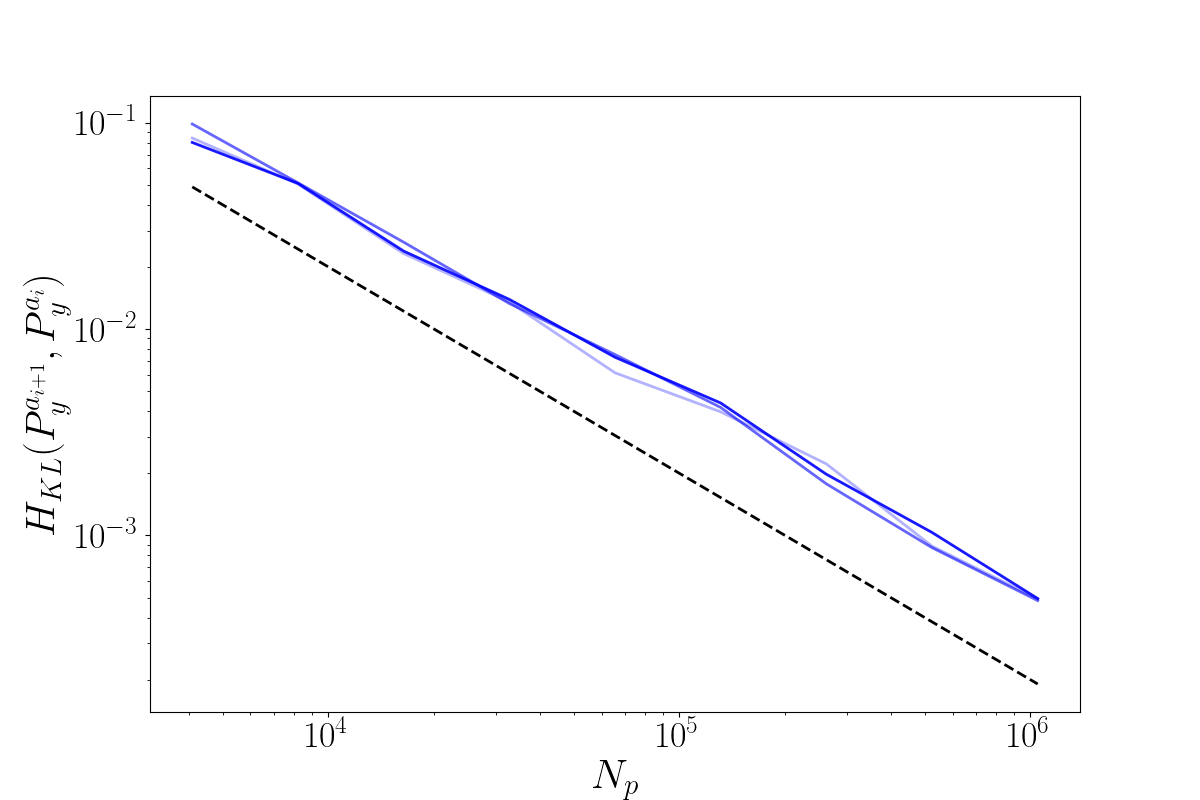}
	\end{subfigure}
	
	\begin{subfigure}[t]{0.45\textwidth}
		\centering
		\includegraphics[width=9cm, center]{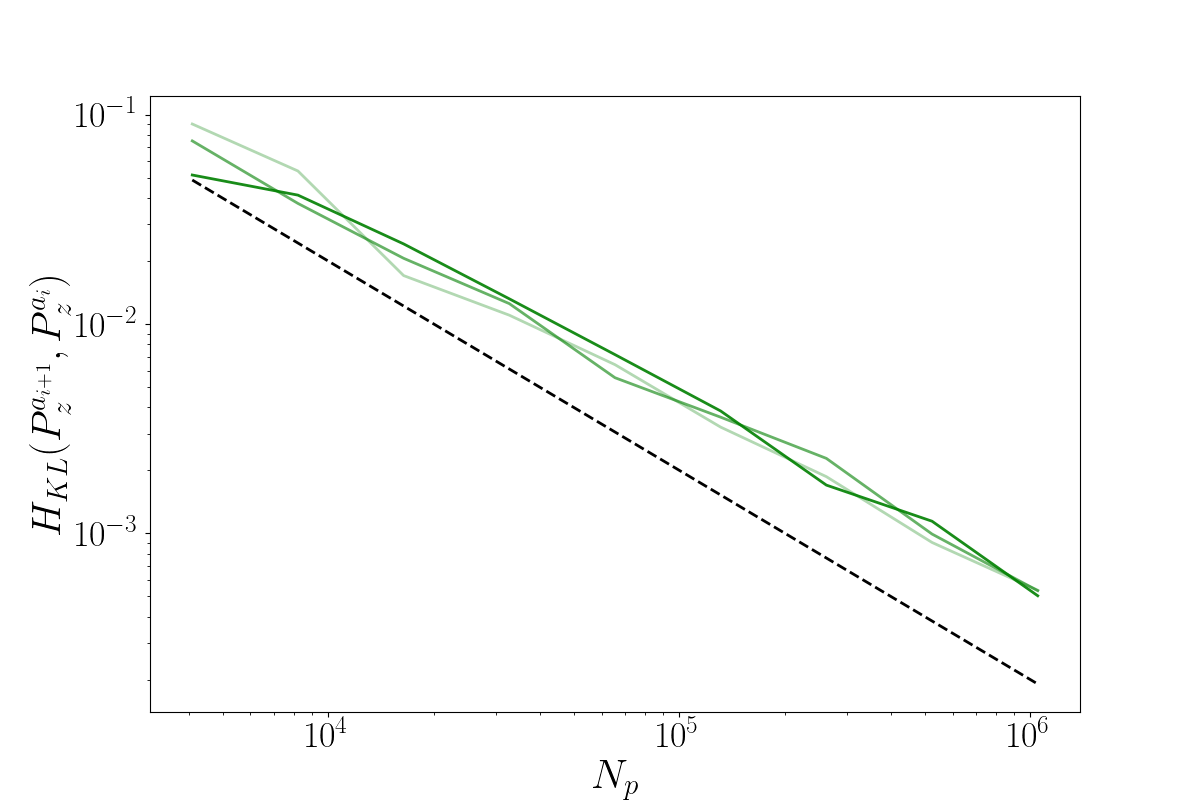}
	\end{subfigure}
	
	\caption{Kullback-Leibler divergence between one-point distributions of increasing number of particles for the WABC model where $h=1/3$ and $N = 14$. All plots show a $N_p^{-1}$ power-law represented by dotted lines.}
	\label{fig:NtrajConv}
\end{figure}

We now evaluate the number of particles necessary to achieve statistical convergence. To that aim, we integrate different numbers of trajectories $N_p$, all emanating from $\bm x_0^*$ using three different time steps: $a_0 = 0.012$, $a_+ = 4 a_0 = 0.048$ and $a_- = \frac{a_0}{4} = 0.003$. We then compute the Kullback-Leibler divergence $H_{KL} (P^{N_p^{i+1}}, P^{N_p^i})$ between distributions of increasing number of particles. We finally compare results between the three time steps to identify if the fluctuations observed in the other figures are due to numerical integration or statistical convergence. Results for the Langevin-WABC setup are presented in Figure \ref{fig:NtrajConv}.

The number of particles does indeed decrease statistical error, at a linear rate. However, the integration constant $a$ has very little effect on those plots as no saturation or change of regime is observed as $a$ increases. The saturation of convergence observed in the Kullback-Leibler divergence study is therefore mainly due to a lack of statistics.

\section{Noise scaling} \label{Ap:Noise}
The noise scaling according to the regularisation appears to be an important parameter to reach spontaneous stochasticity. As illustrated by Bandak et al. \cite{eyink_renormalization_2020} in a simpler model, if the noise is too strong compared to the local dynamics, then the system becomes only driven by the random fluctuations. On the contrary, if the noise is too small compared to the regularisation, it could require to reach very large number of modes $N$ to really observe spontaneous stochasticity. In our more complex example, it is crucial to find an optimal scaling to avoid unnecessary costly computations. We here review the methods we used to determine the noise scaling for both the Cauchy-WABC and the Langevin-WABC cases.

\subsection{The Cauchy-WABC case}
Since the uncertainty on the initial position represents the stochastic regularisation, $\eta_N$ should decrease with the number of modes $N$. An elementary cell in this model is given by its smallest scale $k_N^{-1}$. All particles starting in this cell will have the same initial velocity. The resulting trajectories would inevitably be totally identical for up to a certain time. This phenomenon would ask for longer integration and false durations for the appearance of spontaneous stochasticity. We therefore decided to take an initial ball of a radius of $10$ elementary cells as: $\eta_N = \frac{10}{k_N}$. This allows for a large diversities of initial velocities.

Results obtained in Figure \ref{fig:HKLConvCauchy} show a convergence of statistics at around $N = 14$ which still requires an acceptable large computation time.

\subsection{The Langevin-WABC case}
The noise scaling in this case is not trivial to determine. In particular, the use of a dimensional argument to determine a noise scaling is not optimal. In our trials we noted that such scaling does not show convergence of statistics up to $N = 18$, which is already a computationally intense setup. In order to find a scaling that would allow for a faster convergence, we set a general scaling of the form:
\be
\epsilon_N = b^2 \frac{\omega_N^2}{k_N^{p(h)}}.
\ee
Our goal is to determine the function $p(h)$. To illustrate the process, we here only focus on $h = 1/3$. We use the Kullback-Leibler divergence $H_{KL}$ in order to compare the one-point probability distributions from: Cauchy-WABC $p^C$ and Langevin-WABC $p^L$. We recall that the lowest $H_{KL}(p^C, p^L)$, the closest the two distributions are. As such, we should find the optimal $p(1/3)$ when we find $H_{KL}(p^C, p^L)$ is minimum.

\begin{figure}
	\centering
	\includegraphics[width=12cm]{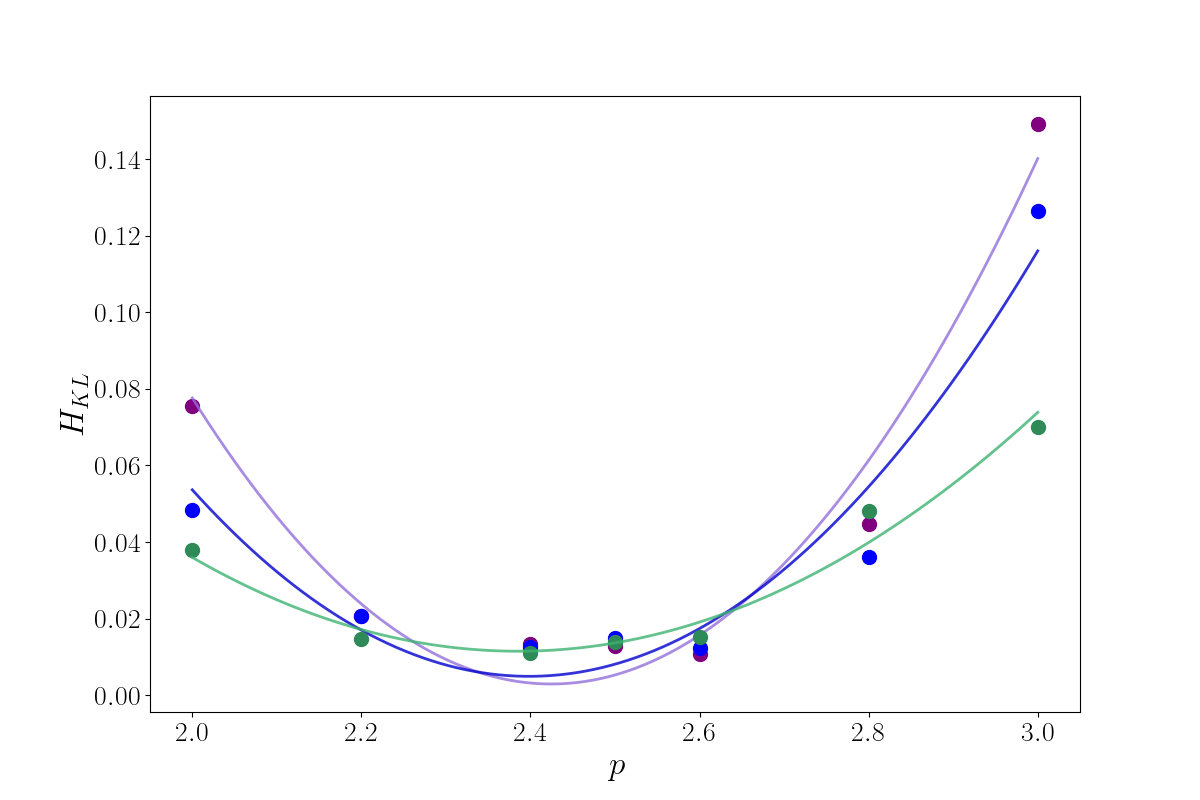}
	\caption{Kullback-Leibler divergence between Langevin and Cauchy one-point statistics' (at time $t = 0.875$) for different exponents $p$ in the case of $N = 16$ and $h=1/3$. Axes are displayed by different colours: $x$-axis is purple, $y$-axis is blue and $z$-axis is green.}
	\label{fig:NoiseOpth3}
\end{figure}

We perform different simulations for $p \in [2, 3]$ and evaluate the one-point probability distributions at time $t = 0.875$. Results are summarised in Figure \ref{fig:NoiseOpth3}.

We notice that, within the error induced by the convergence of statistics (and numerical scheme), the curves are not flat and present some minima for all axes. The initial value $p=3$, guessed from dimensional analysis, appears to be far from the minimum. A simple second order polynomial fit gives us a minimum value for each coordinate, all situated at $p \simeq 2.4$. The non-linearity and intrinsic chaos of such model could explain why dimensional analysis fails here, leading to non-trivial noise scalings.

\end{appendices}

\printbibliography

\end{document}